\documentclass[10pt]{iopart}
\usepackage{graphicx}
\usepackage{multicol}
\usepackage[colorlinks=true,urlcolor=blue,citecolor=blue,linkcolor=blue,breaklinks=true]{hyperref}
\usepackage{color}
\usepackage[dvipsnames]{xcolor}
\usepackage{amstext}

\makeatletter
\renewcommand*{\@textcolor}[3]{%
  \protect\leavevmode
  \begingroup
    \color#1{#2}#3%
  \endgroup
}
\makeatother

\definecolor{plum}{rgb}{0.4, 0.3, 0.4}

\usepackage{iopams}  
\begin{document}

\title[How to read between the lines of electronic spectra: parquet decomposition and fluctuation diagnostics]{How to read between the lines of electronic spectra: the diagnostics of fluctuations in strongly correlated electron systems}

\author{Thomas Sch\"afer$^{a,b}$ and Alessandro Toschi$^{c}$}
\address{$^a$Max Planck Institute for Solid State Research, Heisenbergstra{\ss}e 1, 70569 Stuttgart, Germany}
\address{$^b$Coll{\`e}ge de France, 11 place Marcelin Berthelot, 75005 Paris, France}
\address{$^c$Institute of Solid State Physics, TU Wien, A-1040 Vienna, Austria}
\ead{t.schaefer@fkf.mpg.de}
\vspace{10pt}

\begin{abstract}
While calculations and measurements of single-particle spectral properties often offer the most direct route to study correlated electron systems, the underlying physics may remain quite elusive, if information at higher particle levels is not explicitly included. 
Here, we present a comprehensive overview of the different approaches which have been recently developed and applied to identify the dominant 
two-particle scattering processes controlling the shape of the one-particle spectral functions and, in some cases, of the physical response of the system. 
In particular, we will discuss the underlying general idea, the common threads and the specific peculiarities of all the proposed approaches. While all of them rely on a selective analysis of the Schwinger-Dyson (or the Bethe-Salpeter) equation, the methodological differences originate from the specific two-particle vertex functions to be computed and decomposed.
Finally, we illustrate the potential strength of these methodologies by means of their applications the two-dimensional Hubbard model, and we provide an outlook over the future perspective and developments of this route for understanding the physics of correlated electrons.

\end{abstract}

\ioptwocol

\section{Introduction: Electronic correlations at the one- and two-particle level}

The major challenge to be faced when studying quantum systems with a high degree of correlations between their constituents is the difficulty of disentangling the information of a single particle from the rest of the system. An intuitive illustration of this hurdle can be gained by recalling the {\sl ``Mikado"} game, where the challenge is to pick up single sticks from the stick-jumble of the table, possibly without moving any of the other sticks.
This heuristic picture corresponds to the more formal statement that in a correlated systems even the knowledge of the exact expressions for all one-particle properties, encoded in the one-particle Green functions and self-energy, would not provide a complete information about the underlying physics. 

A good example among many illustrating such an intrinsic feature of electronic correlations is provided by the spectroscopic measurements in iron pnictides \cite{Chubukov2015}, which yield relatively moderate renormalizations of the photoemission spectra  (controlled by the {\sl one}-particle self-energy) and, {\sl at the same time}, major correlation hallmarks in the inelastic neutron scattering spectra \cite{Liu2012,Wang2013,Tam2015} (induced by large {\sl two}-particle vertex corrections \cite{Toschi2012,Watzenboeck2020}) for the same material.

Learning how to look behind the outcome of the most direct one-particle  (e.g. photoemission) spectroscopy experiments represents, thus, a crucial step for fully understanding the physics at work in a correlated material and, thus, for formulating reliable theoretical predictions. 
In this context, novel methodological approaches, specifically designed for these aims, have been introduced  \cite{Rohringer2013a,Gunnarsson2015,Gunnarsson2016,SchaeferThesis,Rohringer2020} by us and other co-workers in the last five years and have been exploited by several groups as a rigorous approach to identify the fundamental physical mechanisms driving the outcome of the one-particle spectroscopy.

\begin{figure*}[t!]
       \centering
       \includegraphics[width=0.95\textwidth]{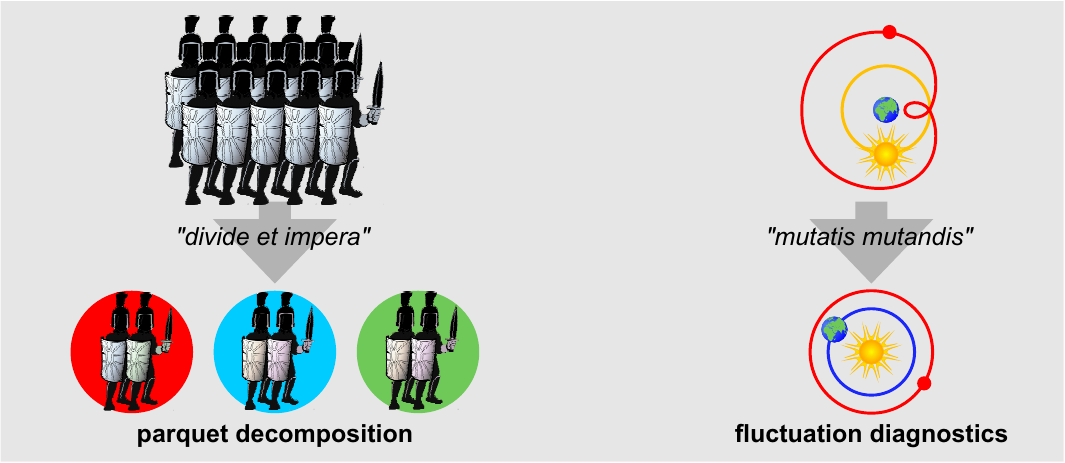}
      \caption{Heuristic illustration of the two main strategies to master the complex theoretical description of electronic correlations. On the left hand panel, inspired by the motto {\it ``divide et impera''} (i.e., ``divide and rule"), one can decompose the full many-body problem in smaller blocks to be dealt with {\sl separately}. As for the specific goal of interpreting the physical mechanisms controlling the one-particle spectral properties, this corresponds to performing a {\sl parquet decomposition} \cite{Gunnarsson2016} of the electronic self-energy, which allows for a precise classification of the scattering processes of different kind. On the right hand panel, consistent with the motto {\it ``mutatis mutandis"} (literally: "having changed what has to be changed"), one can modify the perspective under which the many-electron problem is analyzed, e.g. by changing the reference frame for the description of the physical systems of interest (as a pertinent example \cite{Rohringer2020} we show the representations of the solar systems in different reference frames).
      In our case, this second route corresponds to performing a change of representation of the Schwinger-Dyson equation of motion for the self-energy, in order to exploit the most suited basis to diagnose which fluctuations drives the observed physics - the so called {\sl ``fluctuation diagnostics" }\cite{Gunnarsson2015,Rohringer2020} approach.}
       \label{fig:sketch}
\end{figure*}

\begin{figure*}[t!]
       \centering
       \includegraphics[width=0.85\textwidth]{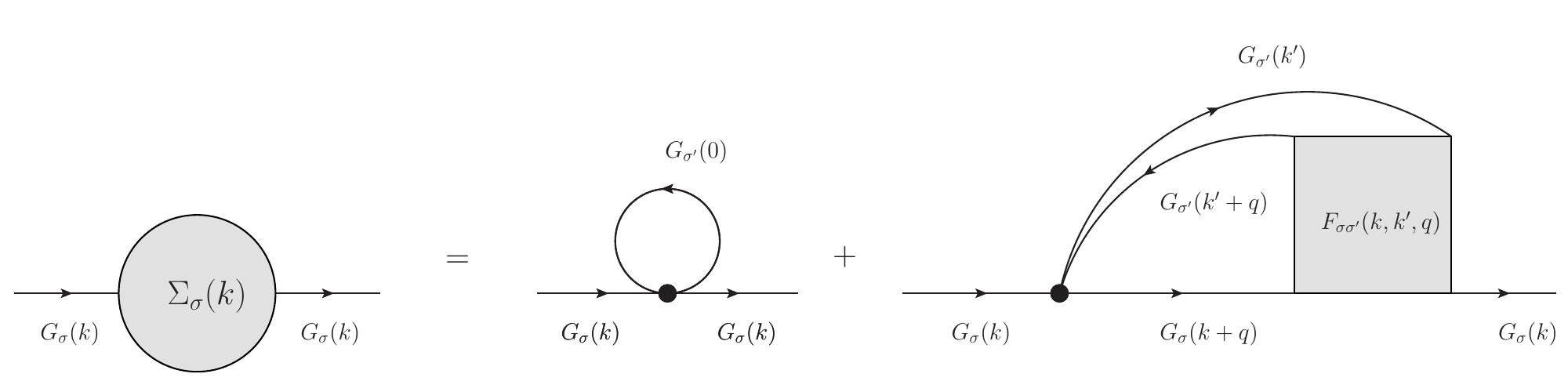}
      \caption{Schwinger-Dyson equation of motion for calculating the (one-particle) self-energy $\Sigma$ from the (two-particle) scattering amplitude $F$ and the Green function $G$.}
      \label{fig:SDE}
\end{figure*}

While the newly introduced theoretical approaches can be classified in two main categories, which we will discuss in detail below, they share a common philosophy and important similarities (see Fig.~\ref{fig:sketch}). They are all based on the idea of identifying the specific nature of the mechanisms driving the observed shape of given one-particle spectral function from a systematic investigation (or ``diagnosis") of the corresponding two-particle scattering processes.
This ``diagnostic" task, which on an experimental level would require to perform several kinds of spectroscopy and transport experiments, can be achieved in the theory by exploiting the fundamental relation which links the calculation of the {\sl one}-particle self-energy $\Sigma$ to that of the {\sl two}-particle full scattering amplitude (or full {\sl two}-particle vertex) $F$, i.e.~the Schwinger-Dyson equation of motion which reads (for the case of a single-orbital model with on-site electrostatic repulsion $U$):
\begin{eqnarray}
\Sigma(k) & = & \frac{Un}{2}  \\
& - & {UT^2}\sum_{k',q} \, F_{\uparrow \downarrow}(k,k',q) \, G(k')G(k'+q)G(k+q),   \nonumber
\label{eqn:SDE}
\end{eqnarray}
and whose Feynman-diagrammatic representation is explicitly reported in Fig.~\ref{fig:SDE}. In this equation, $G$ represents the one-particle Green function, $n$ is the electron density,  $T$ the temperature and $F_{\uparrow\downarrow}$ the full scattering amplitude between electron with opposite oriented spins. With the compact variable $k$, $k'$ and $q$, we compactly indicate a combination of a momentum and (Matsubara) frequencies [e.g. $k=(\mathbf{k},i\omega_n)$]. We adopt here the particle-hole frequency- and momentum convention (see leftmost diagram in Fig.~\ref{fig:pd}).
Eq.~(\ref{eqn:SDE}) illustrates in a rather transparent way, how a significantly large quantity of physical information encoded on the two-particle level ($F$) gets partly averaged away when transferred to the {\sl one}-particle level ($\Sigma$). 

In a nutshell, the way to ``diagnose" the physical mechanisms driving the one-particle spectral properties is to perform  simultaneous electronic calculations at the one- and two-particle level and, then, with these results at hand, to proceed by a systematic inspection of the relation between $\Sigma$ and $F$ as encoded in Eq.~(\ref{eqn:SDE}).

Formally, the novel ``diagnostic"-methodology  can be regarded as an advanced post-processing tool for all many-electron algorithms which allow for a simultaneous calculation --at a reasonable numerical effort-- of $\Sigma$ and $F$. 
The specific procedure through which the vertex $F$ appearing in Eq.~(\ref{eqn:SDE}) is handled and examined defines two main routes for the ``diagnosis" of the spectral properties, which we will analyze in detail in the following two sections together with corresponding, relevant applications realized by us as well as by several other groups around the world. 

\section{A straightforward (but potentially dangerous) route: the parquet decomposition}

The conceptually most natural procedure for evaluating the impact of two-particle fluctuations of different kind on the one-particle spectral properties is to directly {\sl decompose} the full scattering amplitude $F$ appearing in Eq.~(\ref{eqn:SDE}).

\begin{figure*}[t!]
       \centering
       \includegraphics[width=0.95\textwidth]{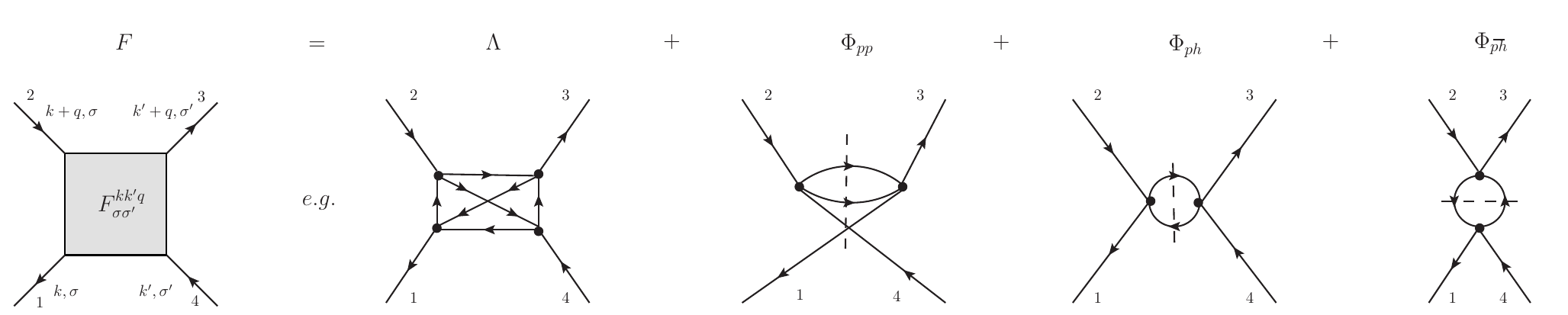}
      \caption{Schematic depiction of the parquet equation which decomposes the full vertex $F$ into a fully irreducible vertex $\Lambda$ and reducible vertices $\Phi_{r}$ and the channels $r=(ph, \overline{ph}, pp)$. The lower row shows examples of diagrams for each of these categories.}
      \label{fig:pd}
\end{figure*}

To this aim, one can exploit the corresponding parquet equation (see Fig.~\ref{fig:pd}):

\begin{equation}
    F=\Lambda+\Phi_{\text{pp}}+\Phi_{\text{ph}}+\Phi_{\overline{\text{ph}}},
\end{equation}
which classifies all two-particle scattering processes contained in $F$ in terms of the two-particle irreducibility\footnote{We recall that a Feynman diagram is defined to be {\sl two}-particle  irreducible (2PI), if it cannot be split into two parts by cutting {\sl two} one-particle Green function lines \cite{Bickers2004,Rohringer2012,Rohringer2018}.} of their Feynman diagrammatic representation in the particle-hole (ph), transverse particle-hole ($\overline{\text{ph}}$) and particle-particle (pp) channel. Reexpressing this equation in the more physical, spin-diagonalized charge (ch) and spin (sp) channels gives:
\begin{eqnarray}
&& F_{\uparrow\downarrow}(k,k'\!,q)\!=\! \textcolor{plum}{\Lambda_{\uparrow\downarrow}(k,k'\!,q)} \!+\! \textcolor{green}{\Phi_{pp,\uparrow\downarrow}(k,k',k\!+\!k'\!+\!q)} \label{eqn:parquet}\\ 
&& +  \textcolor{blue}{\frac{1}{2}\Phi_{ch}(k,k',q)}\textcolor{red}{-\frac{1}{2}\Phi_{sp}(k,k',q) - \Phi_{sp}(k, k+q, k'-k)}. \nonumber
\end{eqnarray}

 The four terms on the r.h.s. of Eq.~(\ref{eqn:parquet}) correspond, thus, to the scattering processes described either by fully two-particle irreducible (2PI)  $\Lambda_{\text{2PI}}$ or to those described by Feynman diagrams which are two-particle reducible in one of the particle-hole [charge ($\Phi_c$) and spin ($\Phi_s$)] or in the particle-particle [pairing ($\Phi_{pp}$)] scattering channel \cite{Bickers2004,Rohringer2012,Rohringer2018}.
By doing so, the self-energy on the l.h.s.~of Eq.~(\ref{eqn:SDE}) gets naturally split into four terms:
\begin{equation}
\Sigma = \textcolor{plum}{{\Sigma}_{\Lambda}}+\textcolor{green}{{\Sigma}_{\text{pp}}}+\textcolor{blue}{{\Sigma}_{\text{ch}}}+\textcolor{red}{{\Sigma}_{\text{sp}}},
\label{eqn:parquetdec}
\end{equation}
allowing for a direct physical interpretation. In fact, the last 
three terms of this {\sl parquet decomposition} of the self-energy identify the contributions to the one-particle spectral properties originating scattering process of a well-defined nature (i.e. arising from charge, spin or pairing fluctuations).

The spirit of this strategy for reading the dominant physical processes behind a given spectral function result is intuitively illustrated on the left hand part of Fig.~\ref{fig:sketch}. 
One uses the parquet equation to split the huge bunch of scattering processes $F$ into smaller parts, each of which allows for a precise classification. This direct 
subdivision procedure ({\it ``divide et impera''}) represents, thus, a key for solving the riddle posed by the physical interpretation of the one-particle spectra.
In this framework, it is easy to figure out that limitations of this approach may originate from the first term in the parquet decomposition. In fact, the 2PI nature of the corresponding processes does not allow for any obvious classification in terms of scattering of a definite kind. Hence, if the contribution stemming from this sub-part of $F$ becomes dominant, which happens, e.g., in the non-perturbative regime (see below), this will significantly challenge the applicability of the decomposition strategy based on Eq.~(\ref{eqn:parquet}).

\subsection{The decomposition of the DMFT and the DCA self-energies from weak-to-strong coupling}
\label{sec:pd_weak_strong}

\begin{figure*}[t!]
       \centering
       \includegraphics[width=0.30\textwidth,angle=-90]{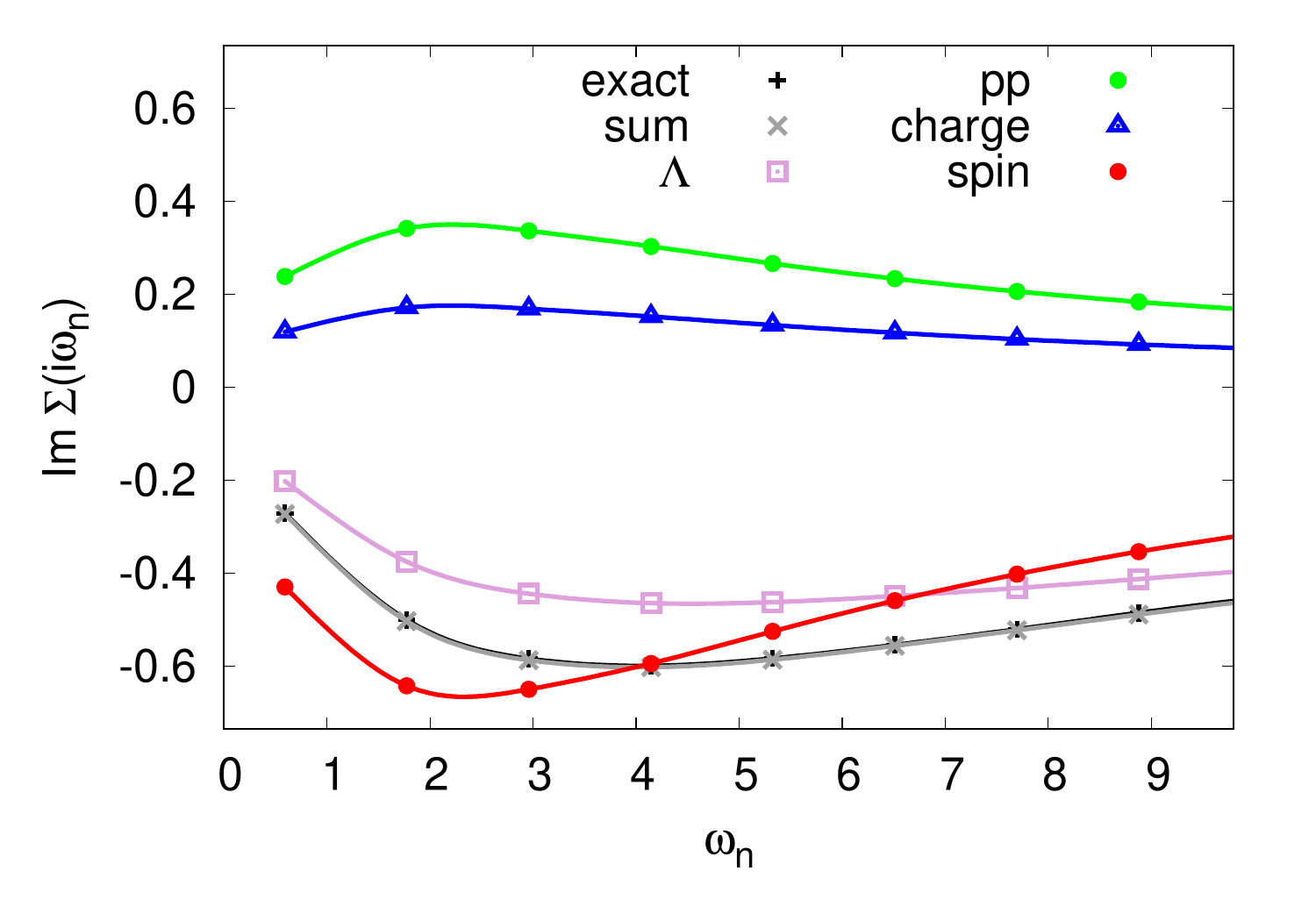}
       \includegraphics[width=0.30\textwidth,angle=-90]{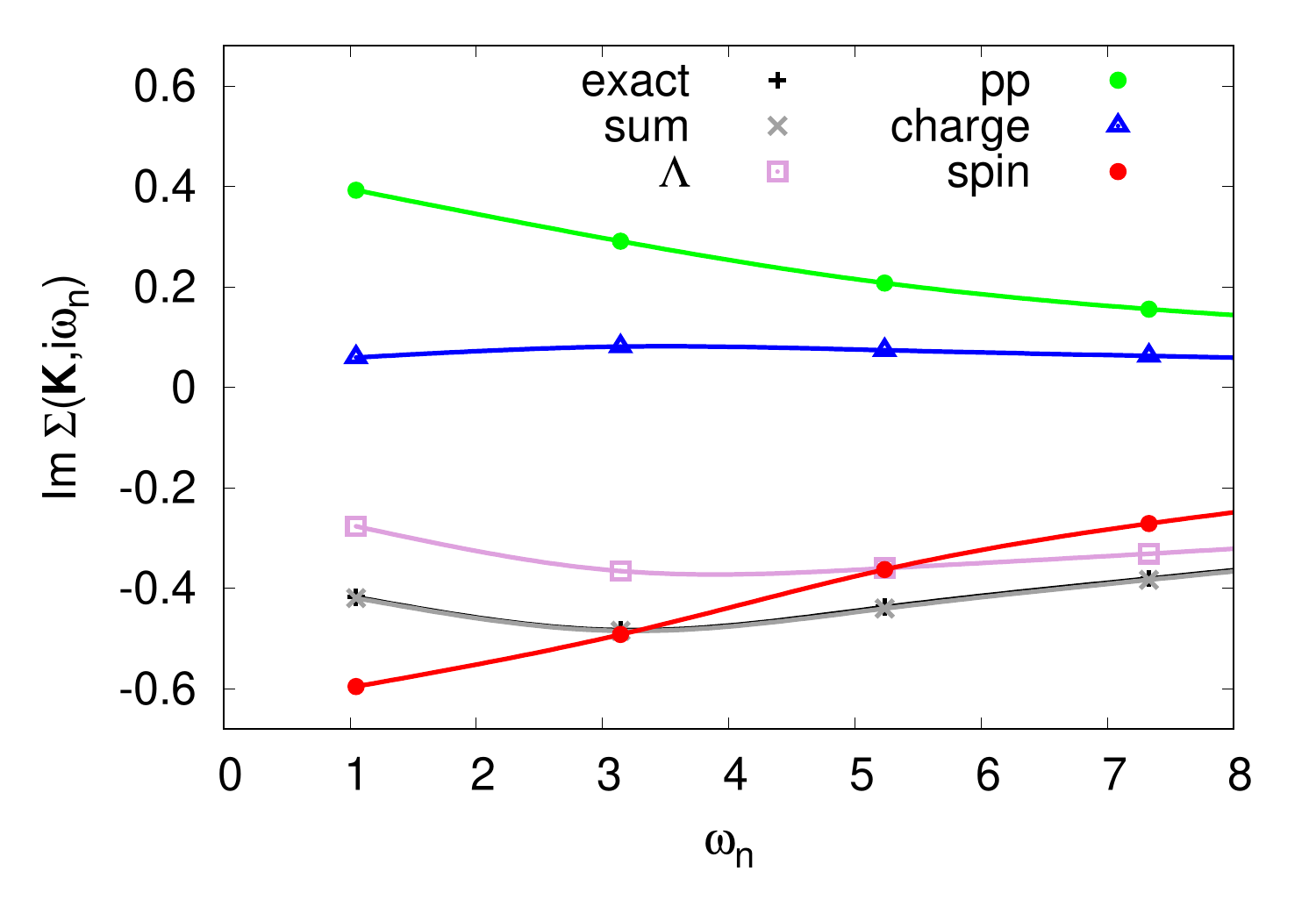}
      \caption{Parquet decomposition of the self-energy in the weak coupling regime. Left: DMFT for the half-filled three dimensional Hubbard model at $U=4.9t$ and $T=0.19t$. Right: DCA for the antinode $\mathbf{K}=(\pi,0)$ in the two-dimensional Hubbard model calculated with $N_c=8$ momentum patches at $U=4t$, $T=0.33t$ and a filling of $n=0.85$. That the sum of the different contributions (``sum") equals the direct calculation of the self-energy (``exact") serves as consistency check (adapted from \cite{Gunnarsson2016}).}
      \label{fig:PDW}
\end{figure*}

\begin{figure*}[t!]
       \centering
       \includegraphics[width=0.30\textwidth,angle=-90]{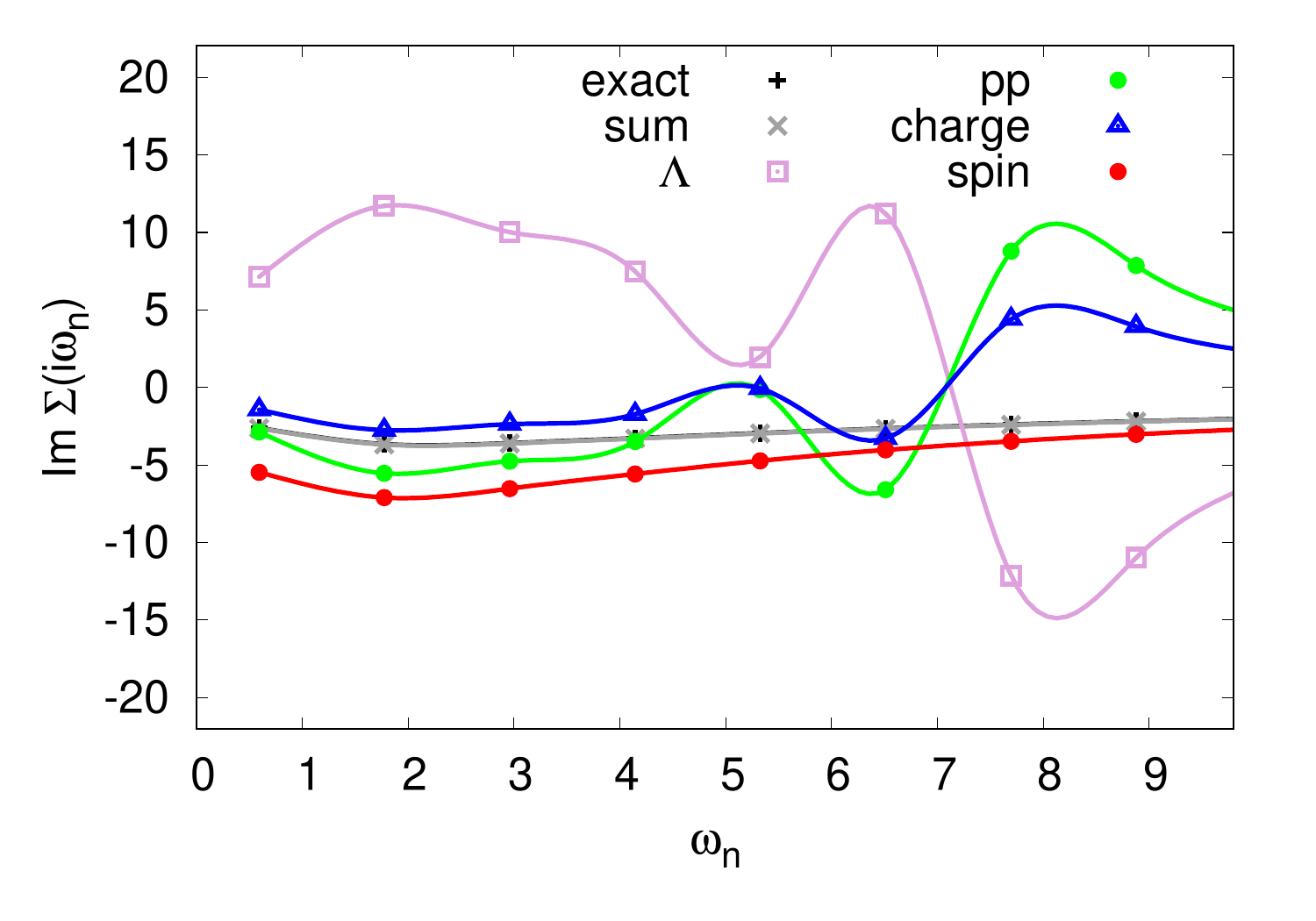}
       \includegraphics[width=0.30\textwidth,angle=-90]{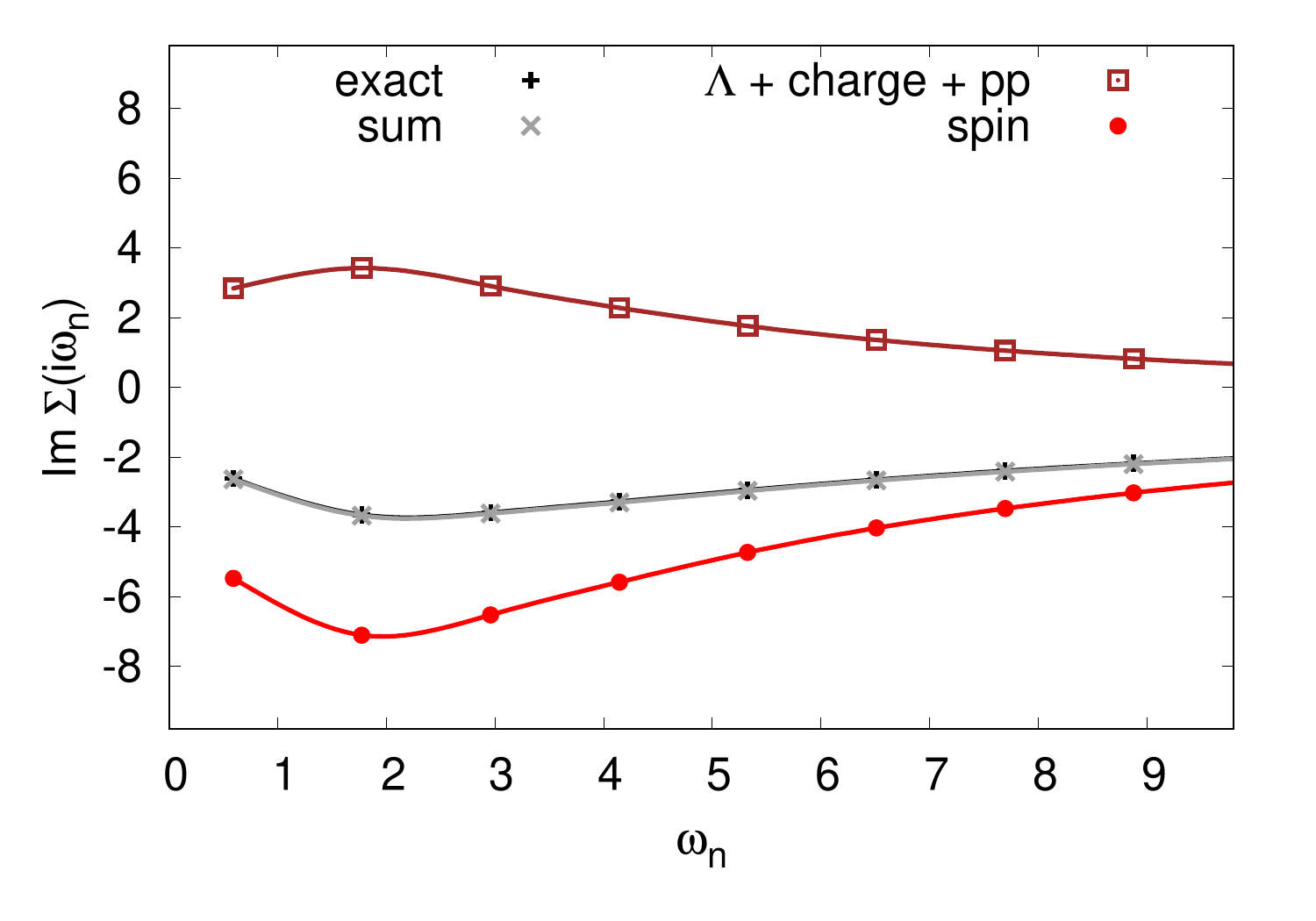}
      \caption{Left: Parquet decomposition of the self-energy in the intermediate coupling regime in DMFT for the three dimensional Hubbard model at half-filling at $T=0.19t$, but for $U=9.8t$. Right: the corresponding Bethe-Salpeter decomposition (adapted from \cite{Gunnarsson2016}).}
      \label{fig:PDS}
\end{figure*}

To illustrate how the parquet decomposition procedure works in practice, it is convenient to start from the systematic analysis of Ref.~\cite{Gunnarsson2016}.
Here the corresponding post-processing procedure is applied to the numerical results of dynamical mean-field theory (DMFT \cite{Metzner1989, Georges1992a, Georges1996}) and one of its cluster extensions, the dynamical cluster approximation (DCA \cite{Maier2005}). Clearly, both algorithms fulfill the prerequisite for a parquet-decomposition as they allow for a simultaneous calculation of self-energy and 2PI vertex functions.
Having all ingredients at hand, the results obtained from the parquet decomposition of DMFT and DCA self-energies of the Hubbard model on the cubic, and respectively, square lattice are reproduced in Figs.~\ref{fig:PDW}-\ref{fig:PDS}. 
In particular, the data shown in Fig.~\ref{fig:PDW} correspond to DCA calculations, where the interaction $U$ is significantly lower than the bandwidth of the square lattice considered ($W\!=\!2D\!=\!8t$). Hence, they can be regarded as representative for the {\sl weak}-coupling regime. 
In both cases the emerging trend is similar: one observes a sizable contribution to the absolute values of Im $\Sigma$ for a momentum (hence: to the overall correlation effects)  originated by scattering processes driven by the spin-fluctuations (in red), which is largely compensated by the screening effects of the fluctuations in the complementary (charge: blue and pairing: green) scattering channels. As a net result the final value of the DMFT and DCA self-energy in the weak-coupling regime reduces to the one driven by the 2PI vertex (which at weak coupling is well approximated by the bare interaction $U$) slightly augmented by an extra contribution originated by the unscreened part of the spin-fluctuations. 
We note that such correction to the contribution of Im $\Sigma$ originating from $\Lambda_\text{2PI}$ reduces to the usual $\frac{U^2}{4i\omega_n}\frac{n}{2}\left(1-\frac{n}{2}\right)$ at high frequencies, where $\Lambda_{\rm 2PI} \rightarrow U$ \cite{Rohringer2012,Rohringer2013a,Thunstrom2018,Chalupa2018} similar \cite{Kunes2011,Chalupa2020} as in perturbation theory. At low-frequency, instead, its role gradually increases by increasing $U$ and, especially in DCA, by reducing $T$.

The physical interpretation emerging from the parquet decomposition in the weak-coupling regime of the Hubbard model is, thus, rather clear: By increasing the interaction strength, the lowest (second) order  contribution to the low-energy spectral properties is progressively enhanced by the scattering of the one added/removed electron onto the spin-fluctuations. 
The latter contribution emerges from a gradually larger imbalance between the (predominant) spin-fluctuation term and the opposite-sign terms describing the screening effects of other channels.

It is interesting to note that beyond an increasingly larger contribution of the spin-fluctuations to the self-energy by enhancing $U$ and lowering $T$, no qualitative change of the physical picture is observed until divergences from $\Lambda_{\rm 2PI}$ are encountered. In this respect we briefly recall that the occurrence of multiple divergences of the 2PI-vertex functions has been proven numerically and/or analytically in all fundamental modellization of correlated electrons, from the DMFT/DCA solution of the Hubbard \cite{Schaefer2013,Gunnarsson2016,Schaefer2016c,Gunnarsson2017,Vucicevic2018,Springer2020} and 
 the Falicov-Kimball \cite{Schaefer2013,Janis2014,Ribic2016} model to the Anderson impurity model (AIM) \cite{Chalupa2018,Chalupa2020} and the Hubbard atom \cite{Schaefer2013,Rohringer2013a,Thunstrom2018}. 
 
 We do not address, here, the general algorithmic challenges \cite{Kozik2015,Stan2015,Gunnarsson2017,Vucicevic2018,Rohringer2018} posed by this rather evident manifestation of a breakdown of the many-electron perturbation expansion at the formal level, nor its recently debated physical interpretation \cite{Kozik2015,Gunnarsson2017,Springer2020,Chalupa2020,Reitner2020}. 
 We observe that since $\Lambda_{\rm 2PI}$ represents the cornerstone of the parquet equation, its divergences are expected to considerably impact all procedures based on a parquet decomposition. Indeed, this is exactly what happens: By increasing $U$ into the intermediate-coupling regime, as it is the case for the DMFT calculations for the Hubbard model on a cubic lattice shown in the left panel of Fig.~\ref{fig:PDS}, it becomes very hard to extract significant information from the parquet decomposition, as the parquet-decomposed contributions to $\Sigma$ start displaying \cite{Gunnarsson2016} wild oscillations in {\sl all} the scattering channels ($ch$, $pp$, as well -of course- $\Lambda$) where divergences of 2PI vertices have been encountered\footnote{We note that, due to the even/odd symmetry characterizing the frequency/momentum structure of the vertex divergences in the particle-hole symmetric situation considered in the DMFT calculations of Ref.~\cite{Gunnarsson2016}, the strong oscillations seen in the Fig.~\ref{fig:PDS} appear only after increasing the interaction beyond the value for which the second vertex divergence, i.e. the first with an {\sl even} frequency/momentum structure, is encountered.}.
 
 \begin{figure*}[t!]
       \centering
       \includegraphics[width=0.90\textwidth]{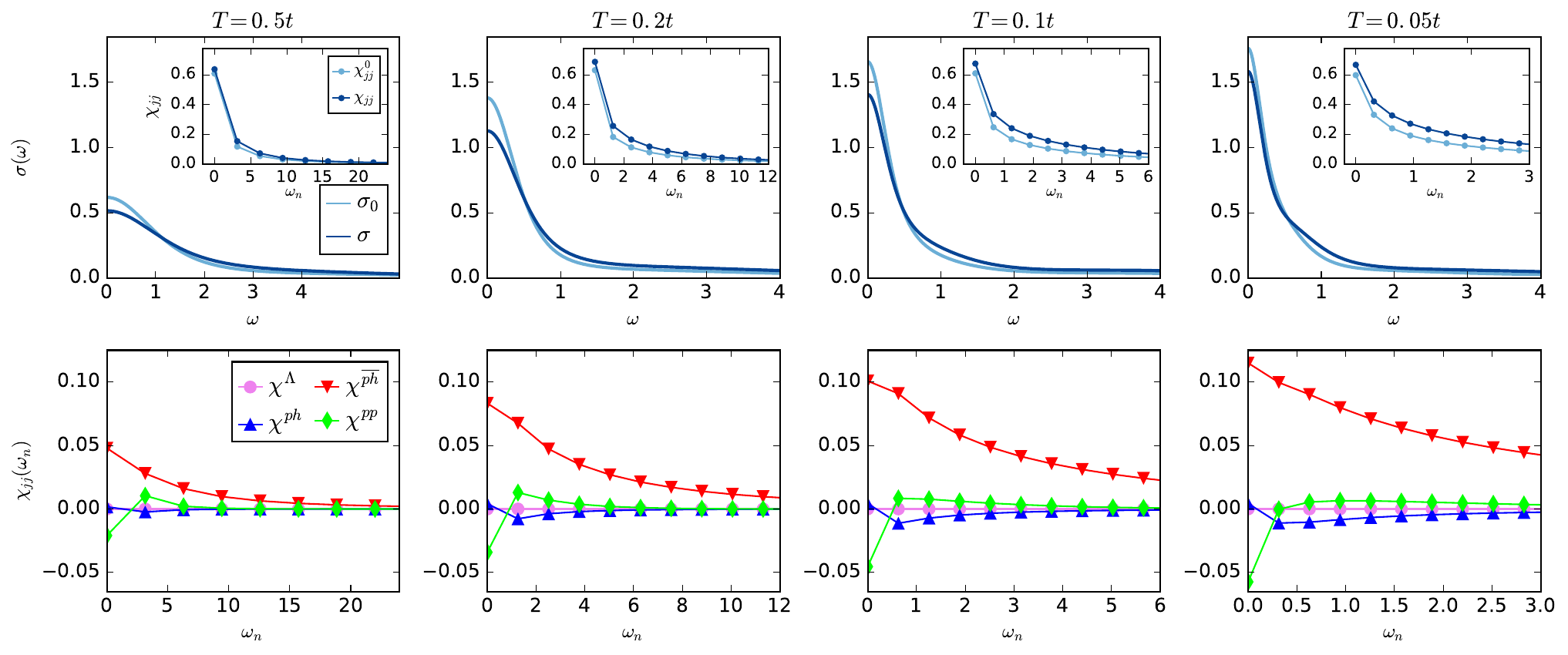}
      \caption{Top panels: Bare-bubble and vertex correction contributions to the optical conductivity computed for the two-dimensional Hubbard model at $U = 4 t$ for different temperatures. Insets: corresponding results for the  current-current response $\chi_{jj}$ as a function of the transfer Matsubara frequency $\omega_n= 2 \pi n T$, with $n \in \mathbb Z$.. Bottom panels: parquet decomposition of $\chi_{jj}$ in terms of the 2PI ($\Lambda$) and the reducible particle-hole, particle-hole transfer/exchange, and particle-particle contributions.
      (adapted from \cite{Kauch2020}).
       \label{fig:piton}}
\end{figure*}
 
 On the other hand, we note that the nasty oscillations of the parquet decomposition  do affect all scattering channels {\sl but} the one which was yielding the largest contribution at lower coupling, namely the spin/magnetic channels. This is highlighted in the right panel of Fig.~\ref{fig:PDS}, where the sum of all strong oscillatory self-energy contributions ($ch$, $pp$, and $\Lambda$) is performed, yielding an overall smooth result. The results of the self-energy can be interpreted in terms of a spin-fluctuation driven physics, where all the other degrees of freedom, however, yield sizable screening contributions. The difficulty we face here, w.r.t. the weak-coupling case of Fig.~\ref{fig:PDW}, is that by means of the parquet decomposition we are no longer able  to disentangle useful information about the precise mechanisms at work in all complementary (screening) channels. 
While this likely reflects a hallmark of the non-perturbative regime, characterized by a strong transfer \cite{Chalupa2020} of information between the dominant and the complementary channels, it poses of course an intrinsic limitation to the full applicability of the parquet-decomposition postprocessing technique in the whole intermediate-to-strong coupling regime of the many-electron systems. Within the heuristic framework of Fig.~\ref{fig:sketch}, one would say that in the strong-coupling regime the division procedure of the  {\sl ``divide et impera"} motto remains possible only for a limited part of the many-electron problem.

Before discussing the impact of such restrictions of the parquet decomposition at the end of the section,  we will illustrate here literature as well as new applications of this post-processing technique in the ``safe" weak-to-intermediate regime, aiming at highlighting the usefulness and versatility of this approach in its region of full applicability. 

\subsection{Versatility of the parquet decomposition}

Hitherto, we have used DCA for illustrating how the parquet decomposition  actually works. However, this should not leave the impression to the reader that the parquet decomposition is specifically designed to analyze  DCA self-energies only. 

On the contrary, the basic nature of the parquet decomposition procedure makes it easy applicable, with minor adjustments, as post-processing tool for a wide range of many-electron calculations of the electronic self-energy. Further,  it can be also exploited in the analysis of other spectral quantities. 

The minimal requirement for the applicability of the parquet decomposition is, as mentioned before, the possibility to compute, separately but on an equal footing,  all the terms of  Eq.~(\ref{eqn:parquetdec}), namely the self-energy for the left-hand side and the 2PI or 2P-reducible vertex functions on the r.h.s..

This requirement is always fulfilled, per construction, in all parquet-based schemes, such as the parquet approximation, dynamical vertex approximation D$\Gamma$A \cite{Toschi2007} (or nanoD$\Gamma$A \cite{Valli2010}) in its most general formulation, and QUADRILEX \cite{Ayral2016}, as well as in the functional renormalization group (fRG)-based methods \cite{Metzner2012}, including the merger between fRG and DMFT coined DMF$^2$RG \cite{Taranto2014,Wentzell2015} and the recently introduced Single Boson Exchange Decomposition SBE \cite{Krien2019c}.

Similar considerations apply if one aims at parquet-decomposing a physical susceptibility $\chi_r$ which can be obtained from the two-particle vertex $F$ through the corresponding Bethe-Salpeter equation (BSE). This schematically reads:
\begin{equation}
    \chi_r = \chi_0 - \chi_0 \Gamma_r \chi = \chi_0 - \chi_0 F \chi_0.
\label{eqn:bse}    
\end{equation}
In practice, after separating the trivial bubble-term part [$\chi_0$ on the r.h.s. of Eq.~(\ref{eqn:bse})], one decomposes the remaining contribution which contains the full vertex $F$ and corresponds to the so-called the vertex correction part of the physical response in terms of 2PI and 2P-reducible vertex functions. This way, one gets the desired four contributions to the physical response, or to be more precise to its vertex correction part. The bubble term, instead does not require any particular diagnostic/decomposition treatment, as it is simply given by the product of two 1P-Green functions.

To demonstrate the versatility of the parquet decomposition, we consider a pertinent example where this scheme has been exploited \cite{Kauch2020} to ``diagnose'' the optical conductivity of the Hubbard model  computed in D$\Gamma$A. The corresponding results are summarized in Fig.~\ref{fig:piton}.
As anticipated above, the first step is to separate the bubble from the vertex-correction term, whose contributions to the optical conductivity $\sigma(\omega)$ (as well as to the current-current response) function are reported in the main panels  (insets) of the top row of Fig.~\ref{fig:piton}.
Secondly, the vertex correction part of current-current response $\chi_{jj}$ has been parquet-decomposed following the procedure summarized above (see Fig.~\ref{fig:piton}, panels at the bottom row) for $U=4t$ (half the bandwidth), at half-filling \cite{Kauch2020}. 
The results of the parquet decomposition shows a clear predominance of the $\overline{ph}$-transverse (i.e., essentially spin/magnetic) fluctuation contribution to the vertex correction part of $\chi_{jj}$, which gradually increases by decreasing temperature.
Similarly as for the self-energy, we see that the contribution of the other scattering channels, i.e. longitudinal $ph$ (which includes the charge fluctuations) and $pp$ goes in the opposite direction, consistent with the interpretation in terms of screening processes. Differently as in the DCA self-energy case above, instead, the parquet decomposition of $\chi_{jj}$ computed in D$\Gamma$A entails {\sl no} contribution arising from the fully irreducible vertex. Such an exact cancellation is a direct consequence of the assumption of pure locality for the fully 2PI vertex $\Lambda_{\rm 2PI}$, on which D$\Gamma$A is based\footnote{Nevertheless, a bare non-local interaction can be included \cite{Kauch2020}, and this adds quite substantial a contribution to the optical conductivity.}. Similar as in DMFT \cite{Georges1996,DelRe2020}, this leads to the vanishing of all corresponding vertex corrections to $\chi_{jj}$ when performing the internal momentum summation, due to the odd symmetry of the current operator. 
Since the assumption of locality of $\Lambda_{\rm 2PI}$ appears a reasonable approximation in large parameter regions of the phase-diagram of the Hubbard model \cite{Maier2006,Rohringer2016}, the contribution $\tilde{\chi}_{jj}^{\Lambda}$ to the current-current response and the optical conductivity is expected to be {\sl marginal} in most cases. 
The major contribution to the vertex corrections arise, thus, by the spatially non-local magnetic fluctuations, partly screened by the charge and the pairing scattering processes. 

Note, however, that this conclusion does {\sl not} apply to the vertex corrections of other response functions such as the charge and the magnetic ones, where the corresponding (charge/spin) operators do not display odd symmetries in momentum space.
In such sectors, the contributions arising from the large, or even diverging fully local 2PI  vertices \cite{Schaefer2013,Schaefer2016c,Springer2020,Chalupa2020}, might trigger relevant physical effects  \cite{Watzenboeck2020,Reitner2020,Chalupa2020}.

\begin{figure}[t!]
       \centering
       \includegraphics[width=0.3\textwidth]{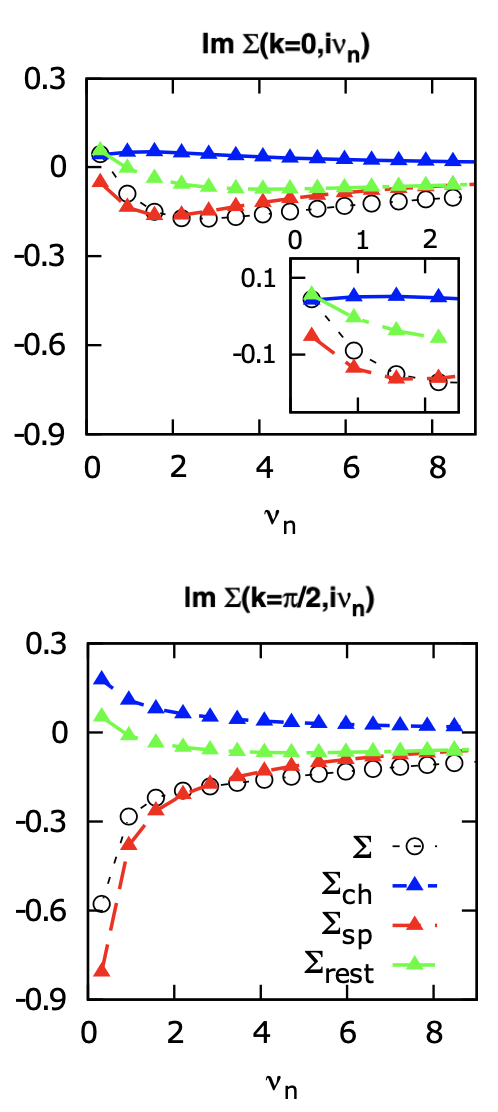}
      \caption{Parquet decomposition of the ladder nanoD$\Gamma$A results for a four-site nano-ring of correlated atoms (adapted from \cite{Valli2015}). Whereas the contributions of the different chaneels are of similar magnitude away from the Fermi surface ($k=0$), the spin contribution dominates at the Fermi surface ($k=\pi/2$).}
      \label{fig:nanodga}
\end{figure}

Finally, it should also be remarked that the versatility of the parquet decomposition allows for its usage even in cases where the parquet structure of the diagrams is not fully resolved, namely as a post-processing of ladder approximations (see also next section).
The obvious limitation here is that only information about the scattering channels whose ladder contribution have been included in the original calculations can be extracted. 
Notwithstanding, the applications of the parquet decomposition has been already proven to be quite useful to interpreting the results of ladder D$\Gamma$A calculations \cite{Valli2015}. 
As an example, let us consider the decomposition of the ladder nano-D$\Gamma$A self-energy $\Sigma({\bf k}, \nu)$ computed \cite{Valli2015} for an isolated nanoring of four correlated atoms (see Fig.~\ref{fig:nanodga}).
In particular, here the parquet decomposition of $\Sigma({\bf k}, \nu)$ has highlighted how the prevalence of magnetic driven contributions was confined to specific momenta, namely to the ones at the Fermi surface.
In the next section, we show how new applications of the parquet decomposition to the ladder D$\Gamma$A calculations allow for a quantitative clarification of the spin fluctuation physics of the unfrustrated two-dimensional Hubbard model.

\subsection{Non-Fermi liquid behavior and different regimes of magnetic correlations at weak coupling}

\begin{figure*}[ht!]
                \includegraphics[width=0.31\textwidth,angle=0]{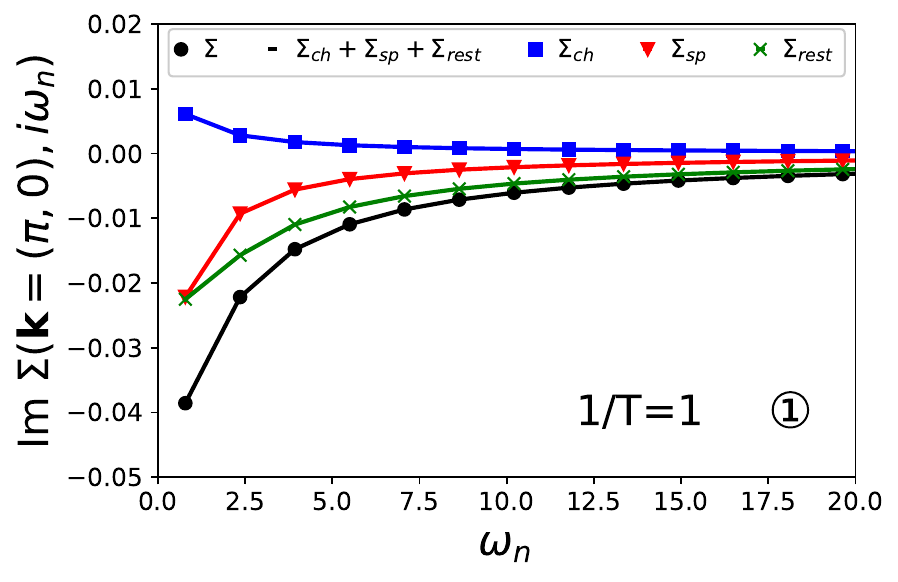}
                \includegraphics[width=0.31\textwidth,angle=0]{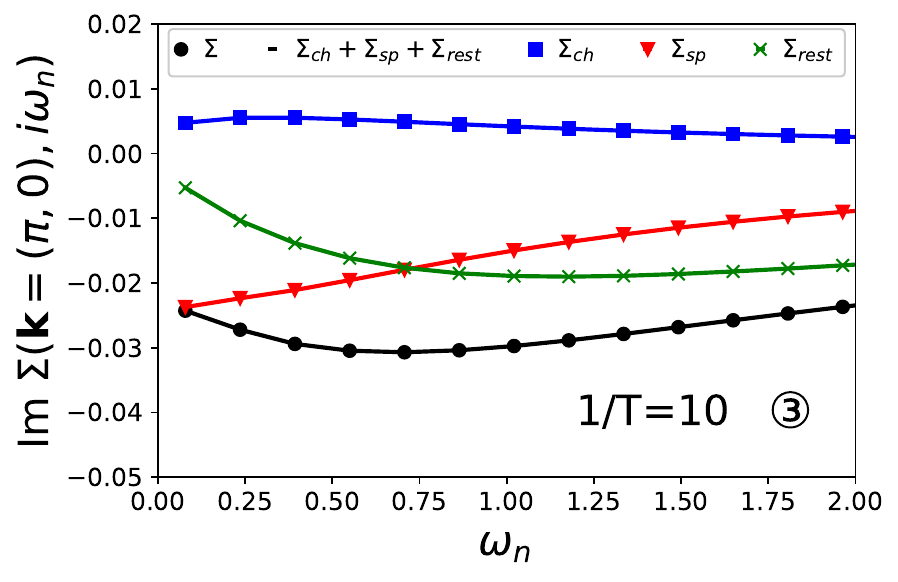}
                \includegraphics[width=0.31\textwidth,angle=0]{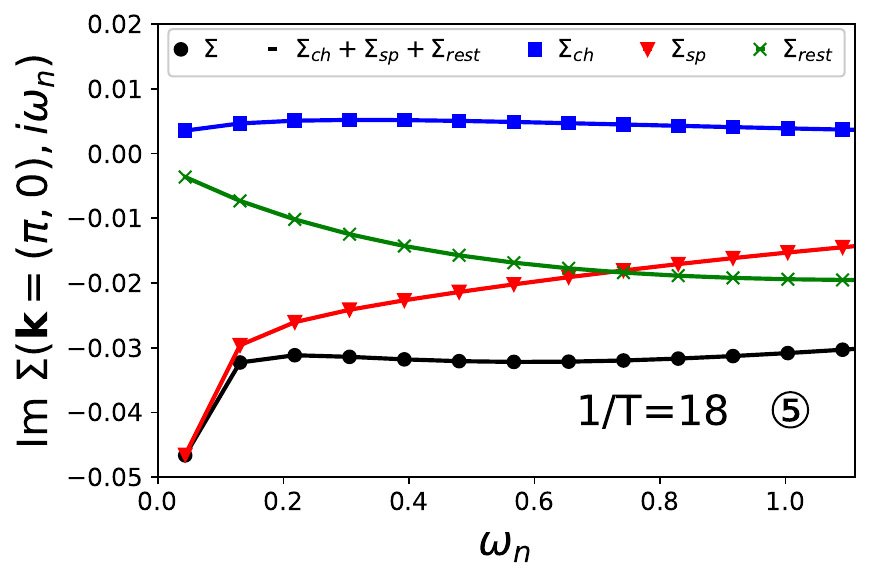}
                \includegraphics[width=0.31\textwidth,angle=0]{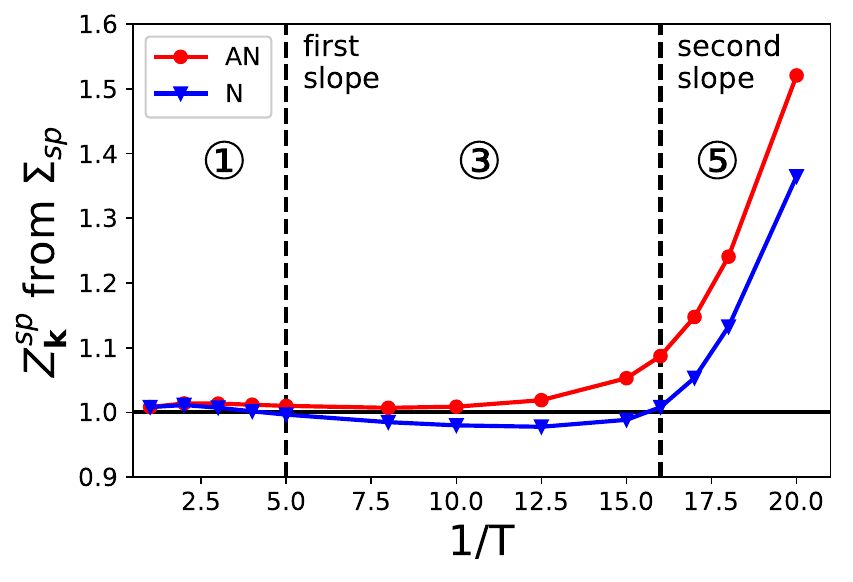}
               \hspace{0.4cm} \includegraphics[width=0.31\textwidth,angle=0]{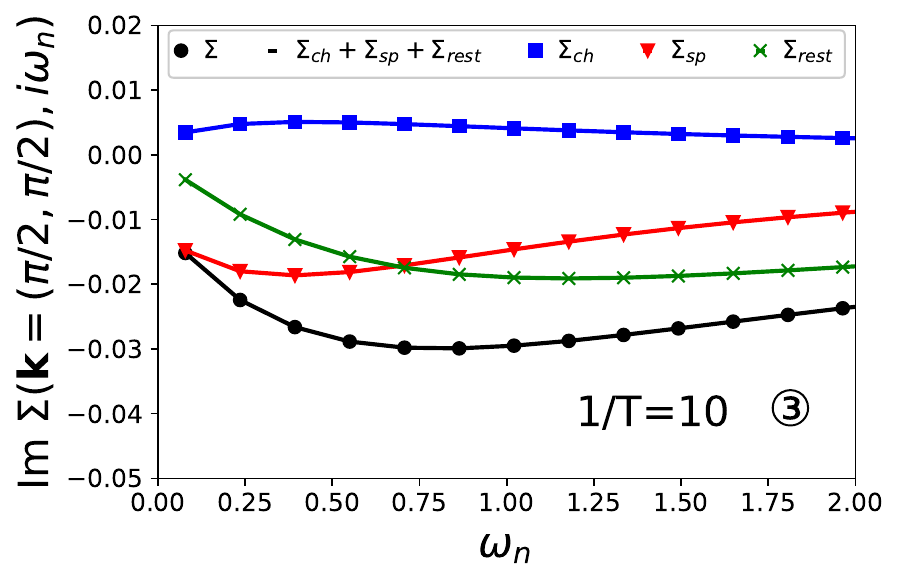}
               \hspace{0.4cm} \includegraphics[width=0.31\textwidth,angle=0]{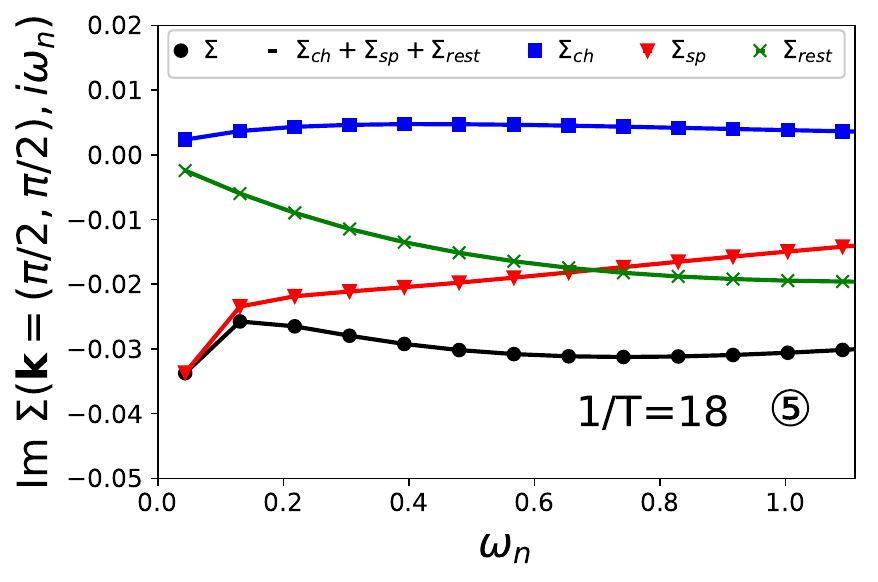}
      \caption{\label{fig:dga_pd} Parquet decomposition of the self-energies obtained by D$\Gamma$A for various temperatures and the quasiparticle weight $Z^\text{sp}_{\mathbf{k}}$ extracted from its spin part. For the highest $T$, only the antinodal point is shown as the nodal one is almost identical to it.}
\end{figure*}

Remarkably, the strategy of {\it``divide et impera"}, founding the base line of the parquet decomposition, can lead to crucial insights even for a case where one could a priori think that the physics is commonly known. To illustrate this, we consider here the case of the two-dimensional Hubbard model on the simple square lattice in the regime of half-filling $n=1$ and weak interaction $U=2t$. This is an arguably "simple" case, also in the sense that there is a plethora or different methods (ranging from "numerically exact" to approximative) available, which can be utilized in this regime of the Hubbard model.

By applying such a multi-method, multi-messenger approach one has been able to nail down its physics and, simultaneously, benchmark each of the methods against numerically exact diagrammatic Monte Carlo (DiagMC) and determinantal (lattice) quantum Monte Carlo (DQMC) \cite{Schaefer2020}. One approximate method which has proven to describe this physics accurately is the dynamical vertex approximation (D$\Gamma$A, here applied in its ladder version \cite{Toschi2007, Katanin2009}), a diagrammatic extension \cite{RMPVertex} of the DMFT \cite{Metzner1989, Georges1992a, Georges1996}.

The black curves with circles markers of Fig.~\ref{fig:dga_pd} show self-energy data computed by D$\Gamma$A as a function of Matsubara frequency $\Sigma(\mathbf{k},i\omega_n)$ for three representative temperature regimes of the Hubbard model at $U=2t$. The first row shows the self-energy for the antinodal point $\mathbf{k}=(\pi,0)$, the second row the nodal one $\mathbf{k}=(\pi/2,\pi/2)$ (at the highest temperature considered, i.e. $1/T=1$, there is no momentum differentiation between nodal and antinodal direction. Hence, the node is not shown for this temperature.). Interestingly, also in this arguably "simple" regime, the model already exhibits an intriguing succession of crossovers as a function of temperature $T$ \cite{Simkovic2020, Kim2020, Schaefer2020, Schaefer2015b, Schaefer2015c}: starting at high $T$, as one cools down, an increase in the coherence of ``quasiparticles" can be observed. This is signalled by a change of slope in the low-frequency behavior of $\Sigma(\mathbf{k},i\omega_n)$ (\textcircled{1} $\rightarrow$ \textcircled{3}), i.e. from a clearly non-Fermi liquid behavior (divergence at low frequencies) to a metallic solution (a Taylor expansion is possible). There is also an intermediate regime present, where only the nodal point shows the onset of coherence (regime \textcircled{2}, not shown). At intermediate temperatures around $1/T=10$ \textcircled{3} the self-energy displays a metallic-like behavior for both node and antinode. Due to the antiferromagnetic ground state, strong magnetic fluctuations set in, when cooling the system further. However, true magnetic ordering at finite $T$ is prohibited by the Mermin-Wagner theorem \cite{Mermin1966, Hohenberg1967} which is preserved by the D$\Gamma$A. As a consequence of these increased fluctuations, Slater paramagnons lead to a divergence of the self-energy at low frequencies (\textcircled{3}-\textcircled{5}) and a (paramagnetic) insulator is established. Please note that also here, a momentum-differentiated regime \textcircled{4} appears, where the nodal point is still "coherent", whereas the antinodal is not (not shown). As demonstrated in Fig.~\ref{fig:dga_pd}, the D$\Gamma$A correctly represents these different regimes, a behavior which has been confirmed by recent DiagMC studies \cite{Simkovic2020, Schaefer2020}.

 We now turn to the parquet decomposition of the D$\Gamma$A results. In the (ladder) version\cite{Katanin2009,Rohringer2016} of the algorithm used, the dominating fluctuation channel (i.e. the magnetic or spin one) was singled out a priori, which, of course, should be directly visible in the parquet decomposition of its self-energy: Focusing on the antinode (first row of Fig.~\ref{fig:dga_pd}) for the moment, one can see that at high temperatures \textcircled{1}, the spin channel contribution $\Sigma_{\text{sp}}$ (red triangles) as well as the "rest" $\Sigma_{\text{rest}}$ (irreducible and particle-particle fluctuations, green crosses) dominate its behavior at low frequencies. The charge channel $\Sigma_{\text{ch}}$ (blue squares) acts as a (small) source of screening. However, as soon as we enter the metallic regime \textcircled{3}, as expected, the spin channel becomes the only dominant contribution. This effect is getting even more pronounced in the low-$T$ insulating regime \textcircled{5}.

\begin{figure}[ht!]
                \includegraphics[width=0.31\textwidth,angle=0]{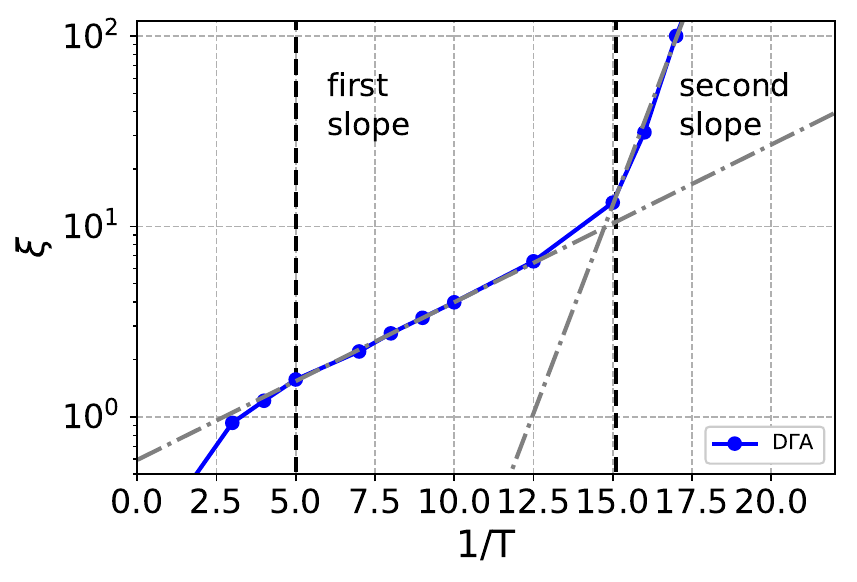}
               
      \caption{\label{fig:dga_xi} Magnetic correlation length calculated by D$\Gamma$A for the half-filled two-dimensional Hubbard model at $U\!=\!2t$.}
\end{figure}
 
 Focusing on the spin contributions $\Sigma_{\text{sp}}$, several points, that cannot be inferred from simply looking at the ``total" self-energy, are noteworthy:
\begin{enumerate}
    \item For the nodal point, the shape of the low frequency (metallic-like vs. insulting/incoherent-like) part of $\Sigma_{\text{sp}}$ is the same as the shape of the complete $\Sigma$. This stands in great contrast to the situation at the antinode: the spin contribution at the antinode {\sl never} becomes metallic-like. Only via the screening of $\Sigma_{\text{ch}}$ and the antagonizing shape of $\Sigma_{\text{rest}}$, metallicity at the antinode can be achieved for the total self-energy.
    \item A straightforward illustration of the former point is shown in the leftmost panel in the second row of Fig.~\ref{fig:dga_pd}: here, the quasiparticle weight extracted from the spin contribution to the self-energy $Z_{\mathbf{k}}^\text{sp}$ is plotted for the two momentum points as a function of (inverse) temperature. In a strict sense, $Z_{\mathbf{k}}^\text{sp}$ is only defined when quasiparticles are present (i.e. in regime \textcircled{3}). However, numerically, one can extract it in every regime from the Matsubara data by
    \begin{equation}
 Z_{\mathbf{k}}=\left[1-\frac{\partial\text{Im }\Sigma(\mathbf{k},i\omega)}{\partial\omega}
 \bigg\vert_{\omega\rightarrow{0}}\right]^{-1}.
 \label{eqn:z}
\end{equation}
 The change of slope from \textcircled{1} $\rightarrow$  \textcircled{3} would be then signalled by $Z_{\mathbf{k}}$ falling $< 1$ (oppositely for the change at low $T$). As discussed before, and as it can be inferred from the figure, the change of slope is {\sl never} happening for $\Sigma_{\text{sp}}$ at the antinode. For the node, $Z^{\text{sp}}_{\mathbf{k}} < 1$ for $16>1/T>5$ and increases above $1$ again for $1/T>16$. So one could argue that the contributions stemming from the spin channel only show a partial metallic-like behavior at the nodal points in a small temperature interval, whereas the antinode is always non-metallic.
 \item Even more interestingly, this regime of partial metallic-like behavior manifests itself also directly at the two-particle level: Fig.~\ref{fig:dga_xi} shows the magnetic correlation length $\xi$ calculated in D$\Gamma$A as a function of inverse temperature. One can see that, starting from a high-temperature mean-field regime (until $1/T \approx 5$), one enters a regime, where the magnetic correlation length is growing exponentially. This is to be expected due to the suppression of magnetic ordering at finite $T$ by fulfilling the Mermin-Wagner theorem in a system with an (antiferro-)magnetic ground state (see, e.g., \cite{Borejsza2003, Vilk1997}). However, quite surprisingly, by cooling the system further, one reaches a {\it second} exponential regime, distinguished by a different prefactor in the exponential. Strikingly, this happens at $1/T=16$, i.e. exactly at the temperature, where both antinodal as well as nodal momentum points (and, in the fully nested system, hence all momentum points) of $\Sigma_{\text{sp}}$ show non-metallic behavior.
\end{enumerate}
This intriguing differentiation of magnetic correlation regimes calls for further investigations  by numerically exact techniques (i.e. going beyond D$\Gamma$A). However, by date, neither DiagMC nor DQMC are able to enter this highly insulating regime.

\begin{figure*}[ht!]
       \centering
       \includegraphics[width=0.90\textwidth]{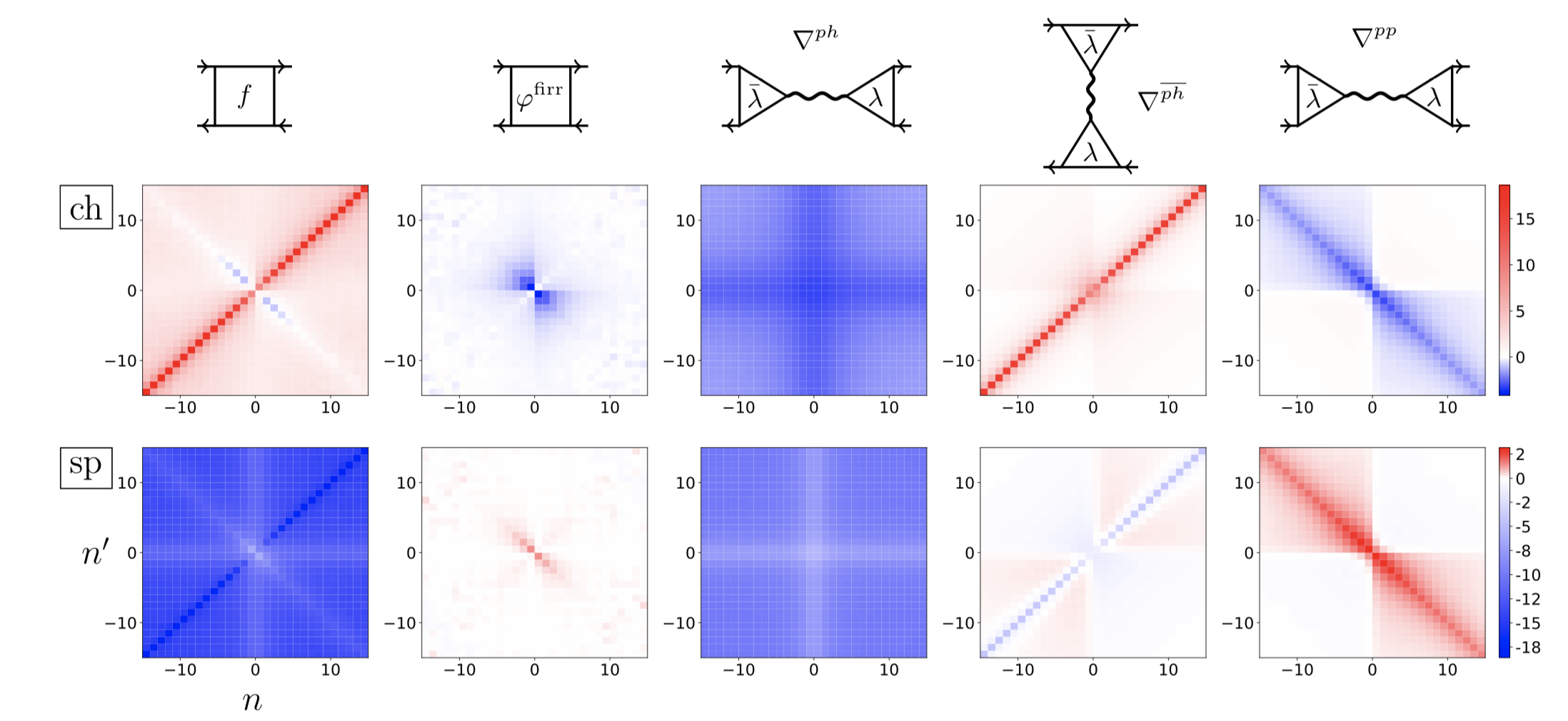}
      \caption{Example of an SBE decomposition: $U$-reducible contributions $\nabla^{r}$ and the irreducible contribution $\varphi^{irr}$ (for an Anderson impurity model corresponding to a self-consistent DMFT solution at $U=4t$) in the charge (upper panel) and spin (lower panel) channel  as a function of the Matsubara frequencies. Also shown is the full vertex $f$ for comparison (taken from Ref.~\cite{Krien2019b}).}
      \label{fig:sbe}
\end{figure*}

\subsection{The challenge of the non-perturbative regime}

As mentioned before in Sec.~\ref{sec:pd_weak_strong}, the applicability of the parquet decomposition faces significant restrictions in the intermediate-to-strong coupling regime, due to the occurrence of multiple divergences \cite{Schaefer2013,Janis2014,Gunnarsson2016,Schaefer2016c,Ribic2016,Vucicevic2018,Chalupa2018,Thunstrom2018,Rohringer2018,Springer2020} of the 2PI vertex functions. 
Of course, the parquet decomposition will remain always applicable to the post-processing of perturbative calculations, such as PA and (multiloop) fRG, where -per construction- singularities of the 2PI vertices cannot appear (except at thermodynamic phase transitions) \cite{Schaefer2016c,Chalupa2018,Rohringer2018}. 
Nonetheless, all the nonperturbative physics intrinsically linked \cite{Gunnarsson2016,Gunnarsson2017,Springer2020,Reitner2020,Chalupa2020} to these 2PI divergences will remain beyond the reach of the parquet decomposition investigation:  The local moment and Kondo-screening regime of impurity models \cite{Chalupa2020},  the Mott-Hubbard MIT \cite{Schaefer2016c,Springer2020} as well as the associated phase-separation instabilities \cite{Nourafkan2019,Reitner2020}, the pseudogap formation as well as the regime of strong AF and RVB correlations \cite{Gunnarsson2016,Gunnarsson2018}, to cite some relevant cases, will be precluded.

Currently, we can envisage two possible ways to circumvent the intrinsic difficulties the parquet decomposition suffers from in the intermediate-to-strong coupling regime. 

The first one requires to recast the parquet formalism in the newly introduced Single Boson Exchange Decomposition (SBE) \cite{Krien2019c}.
This corresponds, essentially, to classifying the two-particle diagrams in terms of their irreducibility/reducibility w.r.t.~a cut of the two-particle interaction instead of two fermionic lines. Within this classification many more diagrams (namely all those corresponding to multi-boson processes) are included in the {\sl irreducible class} of the SBE. 
This inclusion allows for significant cancellations of the divergences displayed by the 2PI and 2P reducible vertices of the usual parquet classification. Indeed, within the SBE, hitherto no vertex singularities have been found \cite{Krien2019c} even up to relatively large values of the electronic interaction.
In practice, one will decompose the full scattering amplitude $F$ in terms of the SBE (instead of the conventional parquet classifications of diagrams).  By means of this procedure which could be referred to as ``SBE decompostion", one will be able to diagnose the self-energy (or the physical response function) by identifying the contributions arising from single-boson-exchange processes in the spin/charge/pairing channel, as well as of a rest part stemming from multi-boson-exchange processes.
While the information encoded in the new classes of diagrams will be different w.r.t.~the usual parquet one, where multiple-boson processes are included in each spin/charge/pairing/$\Lambda$ contributions, the applicability of the SBE-decomposition will not face -a priori- any of the restrictions dictated by the vertex divergences\footnote{In this respect, one could be tempted to draw the conclusion that in large region of the weak-coupling regime of safe applicability of the parquet decomposition, its results would not differ too largely from those computed by means of the SBE one.}. This new kind of decomposition-based post-processing looks particularly promising, because its implementation shares the same versatility of the parquet decomposition. 
For this reason, this approach might be particularly suited to extend the applicability of the diagnostic post-processing to more complex cases, including multi-orbital or long-range-ordered systems.
A first pioneering application of the SBE-decomposition used as post-processing tool has been recently presented in Ref.~\cite{Krien2020c}, see Fig.~\ref{fig:sbe}.

However, also a second, quite elegant, route to perform a ``diagnosis" of the self-energy in the nonperturbative regime, irrespectively of the occurrence of divergences of the 2PI vertices, exists.
This procedure, originally coined as ``fluctuation diagnostics" \cite{Gunnarsson2015,Rohringer2020}, requires a complete change of perspective on the problem, as it will be illustrated in detail in the following section.

\section{Taking a different (and safer) route: fluctuation diagnostics}

The quest of performing a diagnostic of the self-energy in terms of the underlying fluctuations, {\sl without} the restrictions which challenge the parquet-based schemes, has inspired the development of a fully complementary, and quite powerful, post-processing tool.
In fact, as most problems originate from the divergences of the 2PI vertex functions, a  natural way out could be gained by {\sl avoiding} to work at all on the level of 2PI quantities when post-processing the Schwinger-Dyson equation for the self-energy.

The possibility of doing so is offered -in a certain sense- by what it is often considered a problematic issue for the theoretical approaches of the Hubbard model: the Fierz ambiguity. 
In fact, the freedom of recasting the SU(2)-invariant interacting term of the Hubbard model in formally equivalent expressions often leads to somewhat annoying differences 
in the realm of effective fermion-boson and/or purely bosonic approximation-schemes, e.g.~those based on Hubbard-Stratonovic transformations.

Here, instead, the main idea is to reverse the situation and exploit this intrinsic ambiguity to our advantage by deriving formally equivalent representations of the SD-equations which will allow for a selective diagnostic of the underlying fluctuation effects. 

This is, in a nutshell, the essence of the approach which has been coined \cite{Gunnarsson2015} {\sl ``fluctuation diagnostics"} of the self-energy (though more recently such denomination has been also used, somewhat loosely, for any kind of post-processing tool designed for the diagnostics of the self-energy and/or physical response functions).
More specifically, we note that the SD-equation of the self-energy of the Hubbard model in its paramagnetic phase (see Eq.~\ref{eqn:SDE}) can be rewritten in three, formally equivalent ways:
{\small
\begin{eqnarray} 
\Sigma(k)&  &- \frac{Un}{2}  =\\ 
&  &=\,{UT^2}\sum_{k',q} \, F_{\uparrow\downarrow}(k,k'; q) \, G(k')G(k'\! +\!q)G(k \!+ \!q),  \\
       &  &=   - \,
 {UT^2}\sum_{k',q} \, \textcolor{red}{F_{\text{sp}}(k,k'; q)} \, G(k')G(k'\! +\!q)G(k \!+ \!q),  \\
      &  & =   - \, 
 {UT^2}\sum_{k',q} \, \textcolor{blue}{F_{\text{ch}}(k,k'; q)} \, G(k')G(k'\!+\!q)G(k\!+\!q),  \\
            & & =  - \,
  {UT^2}\sum_{k',q} \, \textcolor{green}{F_{\text{pp}}(k,k';q)} \, G(k')G(q\!-\!k')G(q\!-\!k),
  \label{eqn:FD}
\end{eqnarray}}
i.e., in terms of the full scattering amplitude $F$ of the spin, charge or pairing channel\footnote{We recall that the vertex $F_{\uparrow\downarrow}$, due to SU(2)-symmetry and crossing relations, can be re-expressed in terms of the spin $\textcolor{red}{F_{\text{sp}}}=F_{\uparrow\uparrow}-F_{\uparrow\downarrow}$, charge $\textcolor{blue}{F_{\text{ch}}}=F_{\uparrow\uparrow}+F_{\uparrow\downarrow}$ and, via a frequency shift, the particle-particle vertex $\textcolor{green}{F_{\text{pp}}(k,k';q)}=F_{\uparrow\downarrow}(k,k',q-k-k')$.}. 
While the equivalence of the three expressions above after performing all summations on the r.h.s. is just a formal manifestation of the symmetries of the problems (e.g., as the SU(2) one), as it is often the case  with sum-rules \cite{Toschi2005,Toschi2008}, important information can be extracted by considering {\sl partial summations}.

Indeed, by performing all summations but one over the transfer momentum (${\bf q}$) or frequency ($\Omega$), one obtains three (momentum- or frequency-resolved) representations of the electronic self-energy, corresponding to the spin, the charge and the pairing sectors. For instance, for the spin sector, the former would read
\begin{eqnarray} 
\tilde{\Sigma}(k)_\mathbf{Q}&  &- \frac{Un}{2}  =\\ 
       &  &=   - \,
 {UT^2}\sum_{k',i\Omega_n} \, \textcolor{red}{F_{\text{sp}}(k,k'; q)} \, G(k')G(k'\! +\!q)G(k \!+ \!q)\nonumber
  \label{eqn:FD_q}
\end{eqnarray}
while the latter is
\begin{eqnarray} 
\tilde{\Sigma}(k)_{i\Omega_n}&  &- \frac{Un}{2}  =\\ 
       &  &=   - \,
 {UT^2}\sum_{k',\mathbf{Q}} \, \textcolor{red}{F_{\text{sp}}(k,k'; q)} \, G(k')G(k'\! +\!q)G(k \!+ \!q).\nonumber
  \label{eqn:FD_omega}
\end{eqnarray}

While the formal equivalence of these three representations is trivially recovered when performing the missing summations, they outline three complementary description of the same electronic self-energy. 
Heuristically, these representations might recall \cite{Rohringer2020} the different reference systems, through which is possible to describe the movement of the Sun and of the planets in our solar systems (see r.h.s. of Fig.~\ref{fig:sketch}).
In principle, all of them can be used for deriving equivalent descriptions of the solar systems, although some of the reference systems, such as the heliocentric one, are evidently much more suited for getting a transparent description of the underlying physics. 

The very same will happen in the fluctuation diagnostics: If the self-energy under examination is mostly controlled by a well-defined bosonic mode in the spin, charge or pairing scattering channel, its partial summation will be dominated by a specific momentum (or frequency) contribution in the corresponding representation of the SDE.
A strongly-polarized partial-sum in a given representation can be thus regarded as the fluctuation-diagnostic analog of the {\sl heliocentric} representation of the solar-system.

\begin{figure}[t!]
       \centering
       \includegraphics[width=0.40\textwidth,angle=-90]{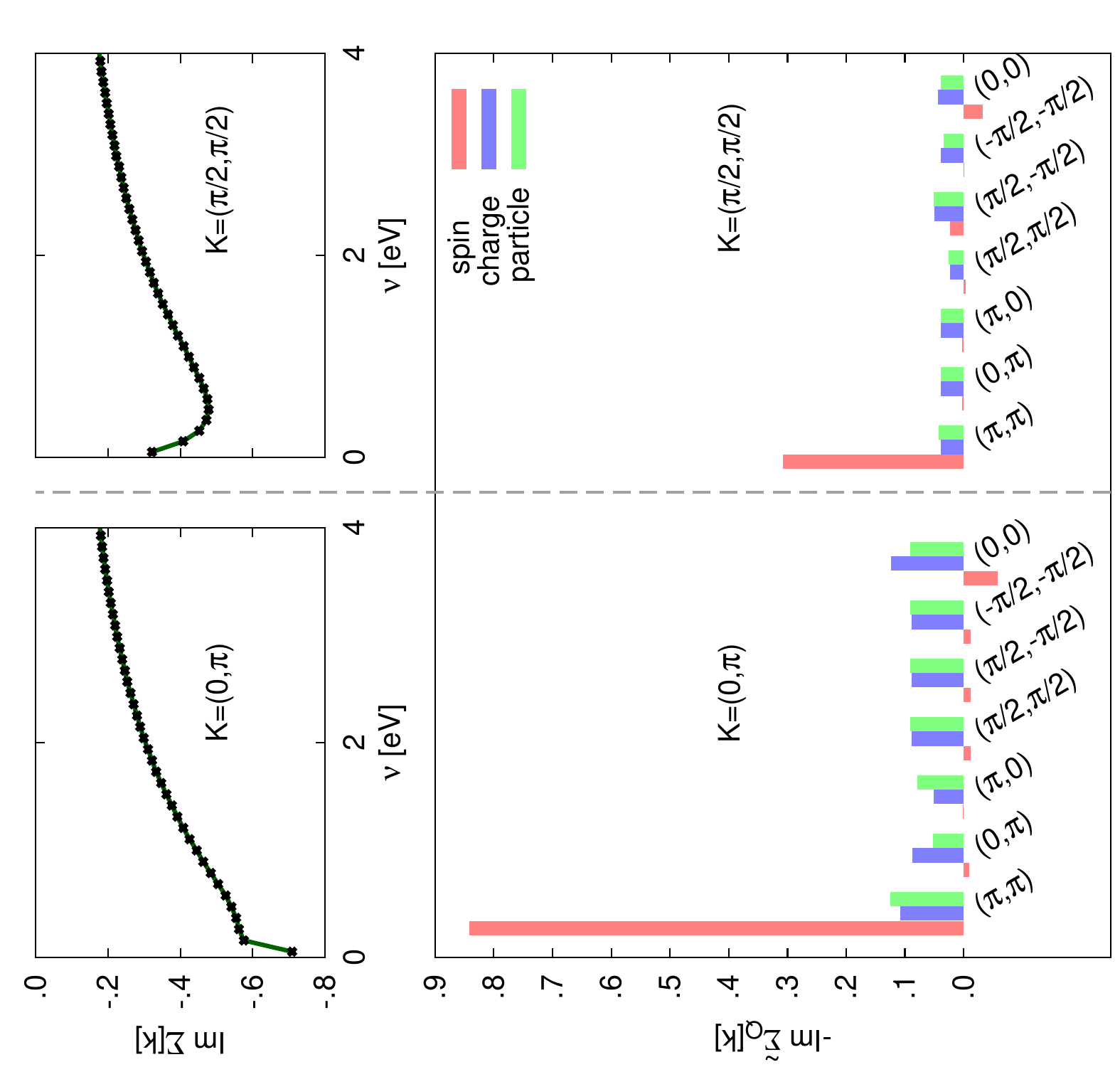}
       \vspace{3mm}
       \includegraphics[width=0.40\textwidth]{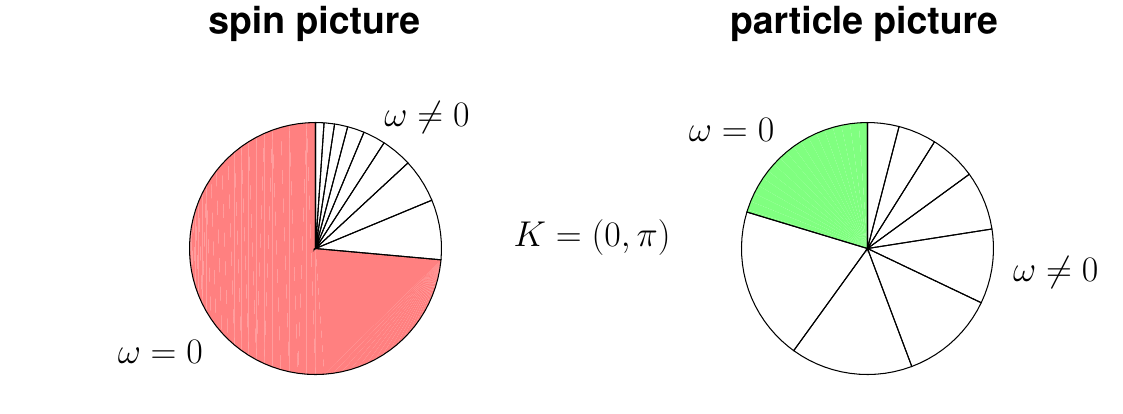}
      \caption{Bosonic fluctuation diagnostics for a DCA calculation in the pseudogap regime of the repulsive Hubbard model ($U=7t$, $t'=-0.15t$, $n=0.94$, $T=0.067t$). Upper row: self-energy at the antinode (left) and node (right). Center row: Momentum-resolved contributions. Lowest row: Frequency-resolved contributions (taken from \cite{Gunnarsson2015}).}
       \label{fig:fd_repulsive}
\end{figure}

Conversely, if we look at the {\sl same} self-energy data in a different representation, we can no longer expect to see a predominance of a specific term: Within this different representation, the fluctuation diagnostic contributions to $\Sigma$ will be distributed rather uniformly as a function of the transfer momentum (or frequency). 
Thus, a featureless shape of the fluctuation diagnostics outcome indicates that the associated representation is {\sl not} particularly suited for a transparent description of the physics underlying the self-energy under investigation. In our heuristic ``astronomical" analogy, this representation would correspond, then, to the {\sl geocentric} reference frame.

\subsection{The fluctuation diagnostics of the DCA self-energies: the origin of the pseudogap}

To guide the reader through the actual results of the fluctuation diagnostics, we start from the 
first applications made in Ref.~\cite{Gunnarsson2015}. 
There, the self-energy  $\Sigma(k)$ of a slightly hole-doped two-dimensional Hubbard model has been computed by means of DCA (with a cluster of 8  discretized momenta {\bf K} in Fourier space). The chosen combination of $U$ and $T$ values considered correspond to the intermediate-to-strong coupling regime of the model. The DCA self-energy (shown in the upper panels of Fig~\ref{fig:fd_repulsive}) displays a significant momentum differentiation between the antinodal [${\bf K}\!=\!(\pi,0)$] and the nodal direction [${\bf K}\! = \!(\frac{\pi}{2},\frac{\pi}{2})$] highly suggestive of pseudogap spectral features.

The fluctuation diagnostics of this DCA self-energy, summarized in the momentum resolved histogram (corresponding to Eq.~\ref{eqn:FD_q}, main panel of Fig.~\ref{fig:fd_repulsive}) and the frequency resolved pie-chart (corresponding to Eq.~\ref{eqn:FD_omega}, lower panel of Fig.~\ref{fig:fd_repulsive}) outline a clear-cut picture of the underlying physics\footnote{The angle of each pie slice corresponds to the fraction of the contribution of the respective Matsubara frequency to the sum over the lowest $n$ (bosonic) Matsubara frequencies ($n=9$ in the given example of Fig.~\ref{fig:fd_repulsive}).}.
Indeed, two of the three representations of the histogram encoding the momentum-resolved partial summation in Eq.~(\ref{eqn:FD})  display a featureless distribution in momentum space, namely the charge one (blue bars) and the pairing one (green bars). 
At the same time, {\it mutatis mutandis}, the fluctuation diagnostic histogram in the spin representation (red bars) is clearly dominated by a specific contribution, namely the one at transferred momentum ${\bf Q} \!=\! (\pi, \pi)$.

Remarkably, such a predominating term is about a factor $two$ larger in the antinodal than in the nodal direction. This result is also supported by the frequency-resolved pie-charts: In the one corresponding to the spin-representation the $\Omega=0$ contribution largely predominates, while in the other two cases a much more uniform frequency-distribution is observed.

\begin{figure}[t!]
       \centering
       \includegraphics[width=0.40\textwidth]{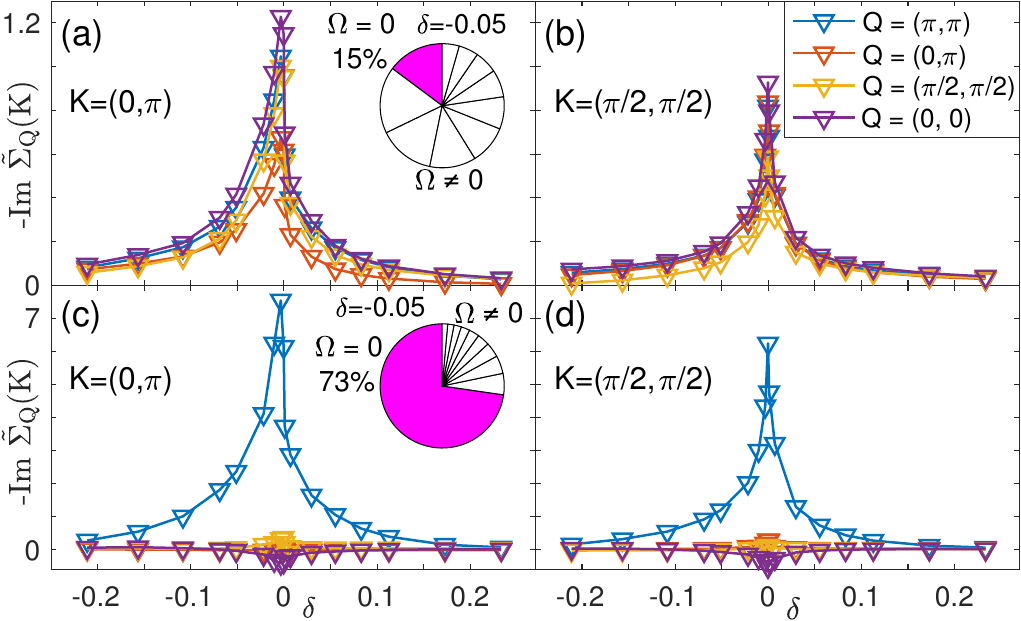}
      \caption{Bosonic fluctuation diagnostics for DCA self-energies at $U=7t$, $t'=-0.15t'$, $n=0.95$ and $T=0.1t$ (taken from \cite{Dong2019}).}
      \label{fig:fddong}
\end{figure}

The spin representation plays, then, the role of the heliocentric reference frame in the solar system example. The underlying  physical interpretation is readily obtained: The relative large and momentum-differentiated values of the DCA self-energy in the pseudogap regime is ascribed almost exclusively to sharply-defined ($\Omega\!=\!0$) AF [${\bf Q} \!=\! (\pi, \pi)$] spin-fluctuations.

Interestingly, a refined analysis of the pairing contributions to Eq.~(\ref{eqn:FD}) has unveiled that, even in the presence of strong $d-$wave pairing fluctuations, these are essentially averaged out by the internal momentum summation over ${\bf K'}$. Hence, the important conclusion of the fluctuations diagnostics introduced in Ref.~\cite{Gunnarsson2015} is not that strong $d-$wave superconducting fluctuations are necessarily absent in the hole-doped Hubbard model, but rather that they do {\sl not} play any decisive role in driving the pseudogap spectral features found in the DCA self-energy.

\begin{figure}[t!]
       \centering
       \includegraphics[width=0.40\textwidth,angle=-90]{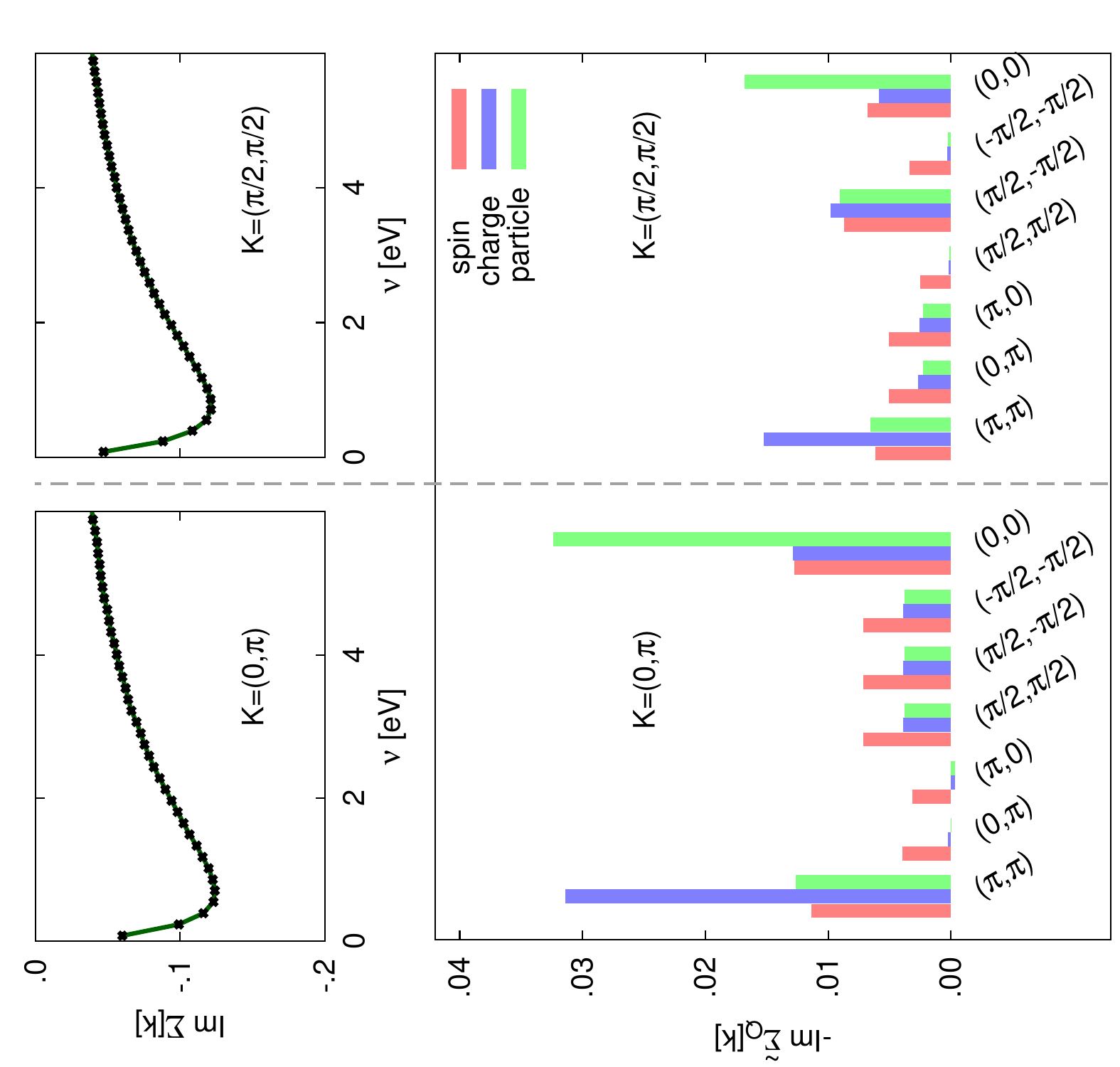}
       \vspace{3mm}
       \includegraphics[width=0.40\textwidth]{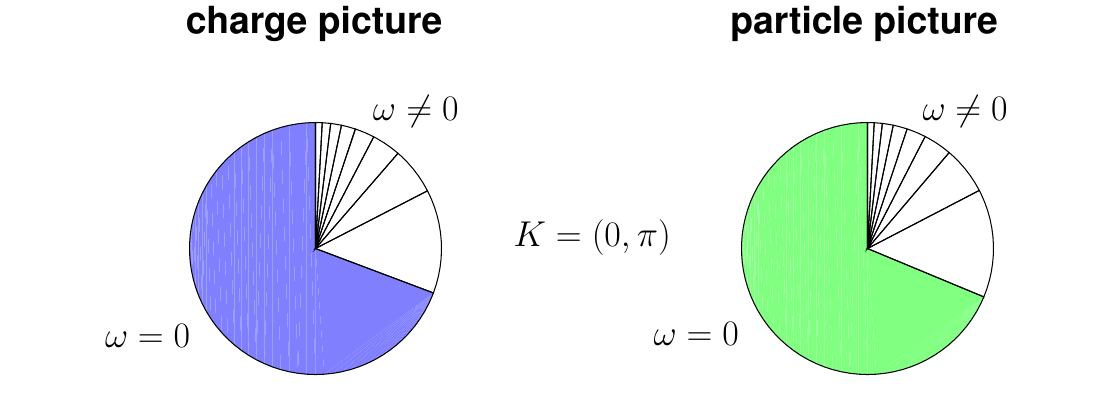}
      \caption{Bosonic fluctuation diagnostics for a DCA calculation for the attractive Hubbard model ($U=-4t$, $n=0.87$, $T=0.1t$). Upper row: self-energy at the antinode (left) and node (right). Center row: Momentum-resolved contributions. Lowest row: Frequency-resolved contributions (taken from \cite{Gunnarsson2015}).}
\end{figure}

A more detailed investigation of the pseudogap regime in DCA has been later performed\cite{Dong2020}, with a particular focus on the role played by the charge fluctuations. These highly precise DCA results will allow for a closer comparison with spectroscopic absorption experiments. At the same time, their fluctuation diagnostic postprocessing, reproduced in Fig.~\ref{fig:fddong}, provides further support to the earlier conclusions of Refs.~\cite{Gunnarsson2015,Rohringer2020}.

The clear-cut result of the pseudogap regime of the repulsive Hubbard model should not leave the reader with the incorrect impression that the predominance of a specific collective-mode can be observed only in one of the three representations. In fact, as it is possible that one gets featureless histograms in all the three representations, it may also happen that two (or more) collective modes are found to yield, simultaneously, predominant contributions in their respective representation.
This is well exemplified by the results obtained \cite{Gunnarsson2015} for the fluctuation diagnostics of the attractive (negative $U$) Hubbard model solved by means of DCA. The corresponding histograms and pie charts identify simultaneously {\sl two} fluctuating modes which contribute (each in its proper representation) to the low-frequency behavior of corresponding electronic self-energy: the $s$-wave superconducting one (${\bf Q}\!=\!0, \Omega\!=\!0$ in the pairing representation, green bars and slices)  and the charge density wave (CDW one (${\bf Q}\!=\!0, \Omega\!=\!0$ in the charge representation, blue bars and slices).
Evidently, this outcome is fully consistent with the physical expectations \cite{Micnas1990,DelRe2019, Tagliavini2016,Springer2020} for the attractive Hubbard model, which provides a further, independent benchmark for the general validity of the fluctuation diagnostics as post-processing tool. 

Let us conclude this section by noting that the fluctuation diagnostics in terms of the bosonic variables (i.e., by performing all internal summation except the one on the transferred momentum/frequency, $q$)  is probably the most natural, but not the only possibility.
As it has been illustrated in Ref.~\cite{Gunnarsson2018}, one can also decide to perform all summations of the SD equation but the one of the remaining fermionic variables $k'$.
This way, one defines a ``fermionic" fluctuation diagnostics which provides complementary, but relevant information w.r.t.~ the most conventional (``bosonic") one. In particular, this fermionic flavour of the fluctuation diagnostics has been applied \cite{Gunnarsson2018} to DCA calculations, aiming at getting more insight on the relation between the RVB correlations and the pseudogap features of spectral function, computed on DCA-clusters between $8$ and $32$ sites.

\subsection{Versatility of the fluctuation diagnostics}

In spite of differences in the implementation, the fluctuation diagnostics shares the quality of being an extremely versatile post-processing tool with the parquet decomposition. 
In fact, the minimal conditions for its applicability to a given many-electron calculation is the possibility of computing one-particle self-energy and two-particle vertex function on an equal-footing, whereas for the latter it will be no longer necessary to extract its irreducible component\footnote{As pointed out in Ref.~\cite{Gunnarsson2018}, the partial momentum/frequency summations to be performed  over bosonic (or fermionic) internal variables of the SD equation might allow for further numerical simplifications, if the values of such partial sums can be directly extracted by the corresponding many-electron algorithm.}.
Hence, the methods for which the fluctuation diagnostics might be exploited as post-processing are essentially the same as for the parquet decomposition, albeit without the restriction of remaining confined to the non-perturbative regime.

\begin{figure}[ht!]
       \centering
       \includegraphics[width=0.40\textwidth]{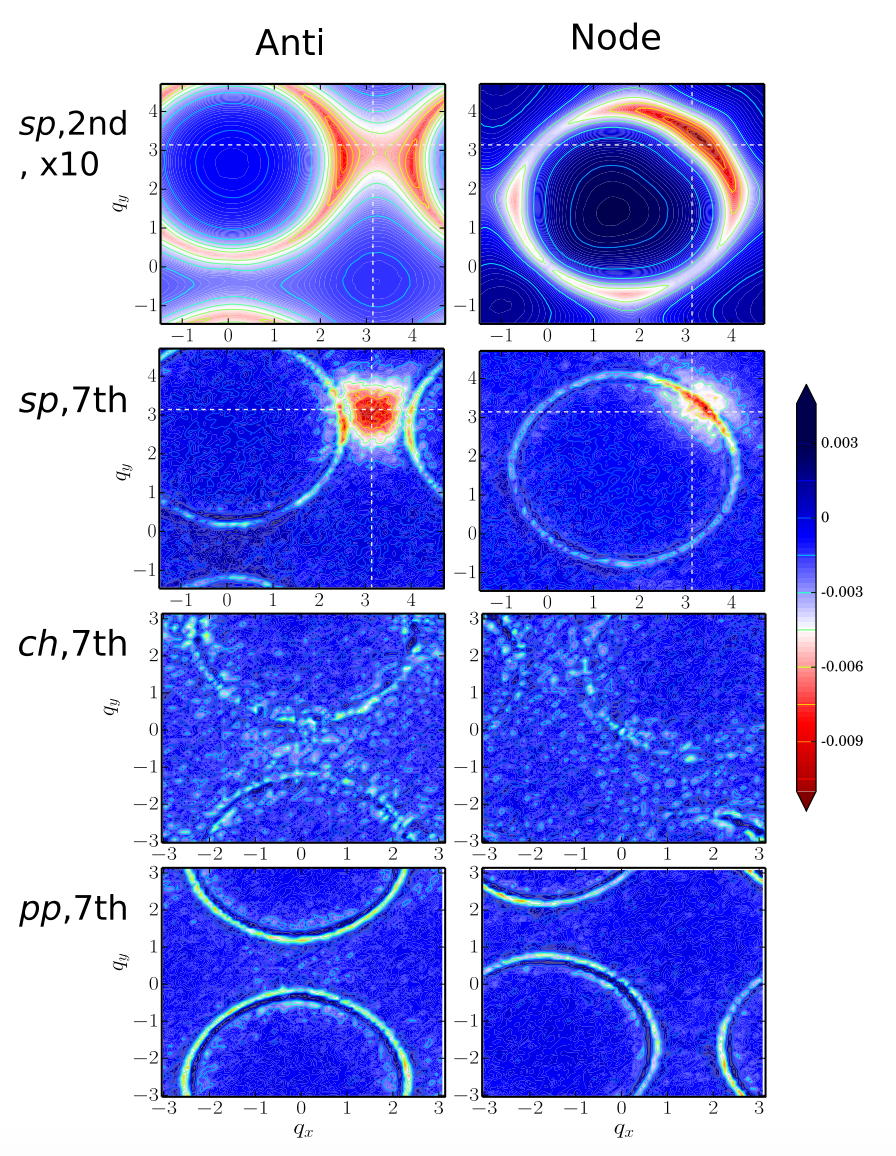}
      \caption{Bosonic fluctuation diagnostics of diagrammatic Monte Carlo results at $U=5.6t$, $t'=-0.3t$, $n=0.94$ and $T=0.2t$ (taken from \cite{Wu2016}).}
      \label{fig:fdwu}
\end{figure}

A relevant example of its broad applicability is represented by the fluctuation diagnostics analysis performed on the Diagrammatic Monte Carlo results of Ref.~\cite{Wu2016}.
Here, the authors have unfolded the partial momentum summation of the SD equation for the self-energy of the hole-doped Hubbard model, and namely for its value a the lowest Matsubara frequency ($\nu \! = \! \pi T)$ and momenta corresponding to the antinodal/nodal direction. The results obtained in \cite{Wu2016} within the three representations of the fluctuation diagnostics are reproduced in Fig.~\ref{fig:fdwu} and correspond to a parameter regime on the verge of that where the pseudogap appears.

We note that, due the high-momentum resolution allowed by this many-electron technique, the partial momentum summations can be presented \cite{Wu2016} in the form of a fully intensity-coloured map over the all Brillouin zone, fully unfurling the potential of the fluctuation diagnostics post-processing approach.
In particular, while the poor significance of the charge and paring modes in shaping the (almost pseudogap-like) self-energy behavior is fully confirmed (as illustrated by the featureless intensity maps of the corresponding representations), new important information was emerging \cite{Wu2016} from the highly momentum-resolved analysis of the spin-representation. 
In fact, by looking at the intensity map of the antinodal and nodal self-energy in this representation, one immediately note the net prevalence of AF fluctuation contribution. However, the comparison between the two shows that, while the fluctuations driving the antinodal self-energy are commensurate (with a maximum at ${\bf q}\!=\!(\pi,\pi)$), the one controlling the nodal self-energy are not, displaying a sizable degree of incommensurability. This refined understanding goes evidently beyond the momentum-grid resolution of the DCA calculations \cite{Gunnarsson2015,Gunnarsson2018} mentioned above. 

Somewhat analogous to the parquet decomposition, we note that also the fluctuation diagnostics can be applied to many-electron calculations which are not fully unbiased w.r.t. the different scattering channels. This is mostly the case of ladder-based approximations\footnote{A similiar procedure might be also useful in the context of phonon calculations \cite{Berges2020}.}. As these approaches imply an {\sl a priori} assumption of the predominance of a specific kind of fluctuations,  the applicability of the fluctuation diagnostics gets restricted, here, to the representations of the channels whose fluctuations are explicitly included through ladder resummations.
The most significant information is extracted, then,  by the inspection of momentum/frequency resolved intensity maps of the selected channels, more than from the comparison of the different representations.

\begin{figure}[t!]
       \centering
       \includegraphics[width=0.40\textwidth]{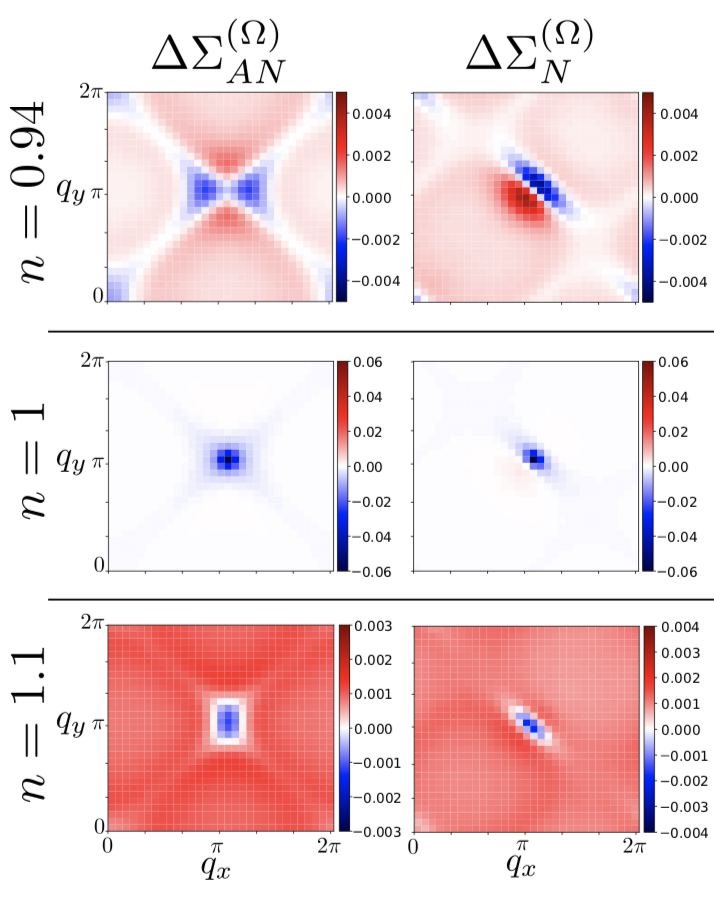}
      \caption{$\Delta\Sigma^{(\Omega)}(\mathbf{q})$ for nodal and antinodal momenta calculated in DF for $U = 5.6t$, $t'=-0.3t$, $T=0.2t$ and $n = 0.94$ (top row), $1$ (middle row) and $1.1$ (bottom row, taken from \cite{Arzhang2020}).}
      \label{fig:fddf}
\end{figure}

As a pertinent example, we report in Fig.~\ref{fig:fddf}, the fluctuation diagnostics recently made in Ref.~\cite{Arzhang2020}. Here, the authors performed the fluctuation diagnostics of the self-energy of a ladder dual-fermion (DF) calculation of the frustrated Hubbard model in the spin-representation. In particular, they have inspected the difference $\Delta \Sigma$ between the self-energy computed at the first $(\nu \! = \! \pi T)$ and the second $(\nu \! = \! 3 \pi T)$ Matsubara frequencies, whose positive/negative sign can be considered, at low-enough temperature, as a reliable indicator of the metallic/non-Fermi-liquid nature of the corresponding spectral function $A_{\bf k}(\omega)$ \cite{Schaefer2015a,Simkovic2020}. 
The data reported in Fig.~\ref{fig:fddf}, which refer to ladder DF calculations performed at different doping levels, show \cite{Arzhang2020} that the pseudogap features of the hole-doped solution (where one finds $\Delta \Sigma \! > \! 0$, i.e. a metal, at the node, and $\Delta \Sigma \! < \! 0$, i.e. a non-Fermi liquid, at the antinode), are built up by qualitatively different partial summations in the spin picture: While at the node, one finds large contributions of opposite sign emerging from AF fluctuations which largely elide themselves, at the antinode the underlying AF-fluctuations act rather synergetically in suppressing the Fermi-liquid coherence. 
It is also interesting to observe that the fluctuation diagnostics looks rather different in the electron-doped case considered ($n\!=\!1.1$, bottom panels of Fig.~\ref{fig:fddf}) and that, there, the spin-representation appears less suited for a complete understanding of the spectral properties, being associated to a more uniformly distributed intensity maps, at a quantitative level.

All of the examples of this section highlight that significant insight into the physics of strongly correlated electron systems can be gained when fluctuation diagnostic approaches are used for post-processing two-particle quantities which are the result of channel-unbiased techniques to the problem at hand. To close this section on the fluctuation diagnostics, in the last subsection, we will ``turn around" this strategy, i.e. we will demonstrate that the fluctuation diagnostics can indicate which method (or its parametrization) is well suited in specific parameter regimes.

\subsection{Estimation of the Fierz parameter in combined
bosonic-fermionic descriptions of the Hubbard interaction}

A common thread of cutting-edge many-body methods is the great amount of simplification that arises when dominating physical channels can be determined a priori. This is true for (purely fermionic) frameworks, eg.~in the simplification obtained by singling out the dominating fluctuations in the parquet framework. The latter  can (under these circumstances) be simplified to a much easier (and numerically much more light-weighed) ladder/Bethe-Salpeter treatment.

In this section we want to highlight the potential of the fluctuation diagnostics with the example of a diagrammatic extension of DMFT, the triply irreducible local expansion (TRILEX, \cite{Ayral2015, Ayral2016a, Ayral2016}). This theory is formulated in a mixed bosonic-fermionic language, where the fermions constituting the system are coupled via bosonic fields and the electron-boson coupling vertex is approximated by a fully local quantity. The formalism can be applied to the Hubbard model by decoupling the electron-electron interaction, e.g. via (up to a density term)
\begin{equation}
    Un_{\uparrow}Un_{\downarrow}=\frac{1}{2}U_{\text{ch}}nn+\frac{1}{2}U_{\text{sp}}\vec{s}\vec{s},
    \label{eqn:decoupling}
\end{equation}
where $U$ is the Hubbard interaction, $n$ are the density operators, $\vec{s}$ are the (Heisenberg) spin operators and $U_{\text{ch/sp}}$ are the channel-dependent (constant) bare couplings of bosons with fermions. These couplings have to fulfill the following relation:
\begin{equation}
    U=U_{\text{ch}}-3U_{\text{sp}},
\end{equation}
leading to
\begin{eqnarray}
    U_{\text{ch}}&=&(3\alpha-1)U\\
    U_{\text{sp}}&=&(\alpha - 2/3)U,\nonumber
\end{eqnarray}
which makes the choice of the couplings depending on an ambiguous number $\alpha$ (Fierz ambiguity \cite{Ayral2016, Ayral2017}). In the following we will see, how fluctuation diagnostic techniques can help to achieve a good choice of the Fierz parameter $\alpha$.\\

\begin{figure}[t!]
       \centering
       \includegraphics[width=0.40\textwidth]{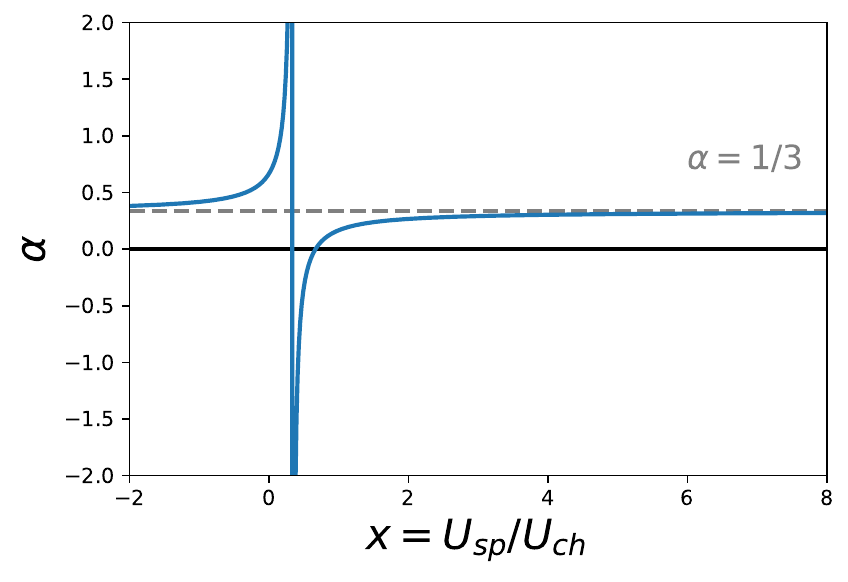}
      \caption{\label{fig:fierz} The Fierz parameter $\alpha$ within the Heisenberg decoupling expressed as the ratio of the spin and charge coupling $x$.}
\end{figure}

\noindent
First, we define the ratio
\begin{equation}
    x \equiv U_{\text{sp}} / U_{\text{ch}}.
    \label{eqn:x}
\end{equation}
Fig.~\ref{fig:fierz} shows the value of the Fierz parameters $\alpha$ depending on this ratio $x$. For estimating the Fierz parameter in the pseudogap regime of the Hubbard model, we resort to the fluctuation diagnostics presented in Fig.~\ref{fig:fd_repulsive}: If we take the ratio of the self-energies in the spin representation [at $\mathbf{Q}=(\pi,\pi)$] and the charge representation [at $\mathbf{Q}=(0,0)$] for the lowest Matsubara frequency at, e.g., the antinode as a measure for their relative strengths, we arrive at:
\begin{equation}
    \frac{\text{Im }\tilde{\Sigma}_{\text{sp}}(\mathbf{k}=(\pi,0),i\omega_0)}{\text{Im }\tilde{\Sigma}_{\text{ch}}(\mathbf{k}=(\pi,0),i\omega_0)} \approx 4.
\end{equation}
Taking this value as a proxy for the actual coupling ratios $x$, via Fig.~\ref{fig:fierz} we arrive at an estimated $\alpha \approx 1/3$. Interestingly, this represents a decoupling solely in the spin channel. If the spin fluctuations would dominate even more, due to the asymptotics of this curve, the estimated parameter would still be $\alpha \approx 1/3$. We can put these assumption to solid testing grounds by applying the procedure to a case where we know the dominant fluctuation channel, i.e. the Hubbard model at half-filling and weak coupling $U=2t$. There, due to its antiferromagnetic ground state, large spin fluctuations are present even at finite temperatures, so that the ratio of spin and charge fluctuations $x$ should indeed fall into the asymptotic regime of Fig.~\ref{fig:fierz}.

\begin{figure}[t!]
       \centering
       \includegraphics[width=0.40\textwidth]{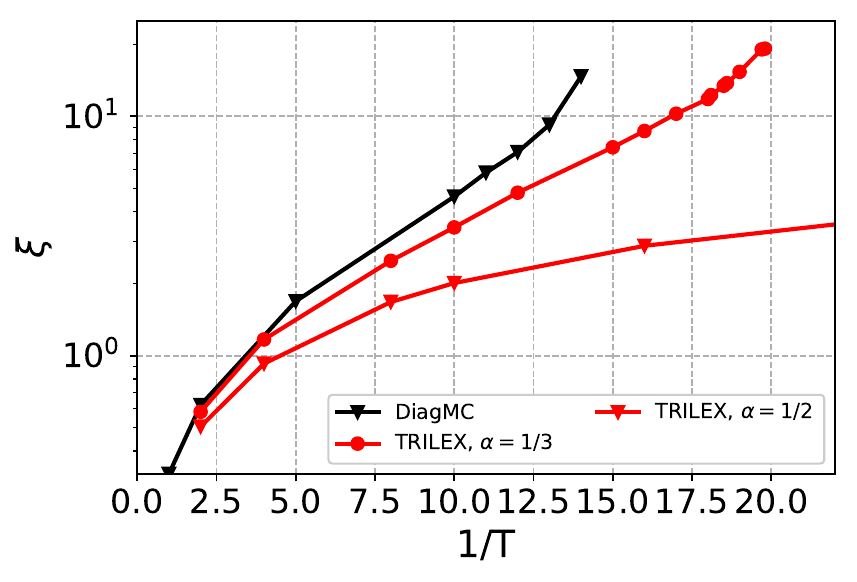}
      \caption{\label{fig:xi_TRILEX} Magnetic correlation length $\xi$ calculated within the TRILEX approximation, calculated with two different Fierz parameters: $\alpha=1/3$ (red circles) and $\alpha=1/2$ (red triangles). It is compare to a diagrammatic Monte Carlo benchmark (black triangles, data taken from \cite{Schaefer2020}).}
\end{figure}

In Fig.~\ref{fig:xi_TRILEX} we show TRILEX data for the magnetic correlation length $\xi$ calculated for two different Fierz decouplings (red triangles: $\alpha=1/2$, red circles: $\alpha=1/3$) compared to benchmark data from DiagMC (taken from \cite{Schaefer2020}). One can see that the $\alpha=1/3$ data much better resemble the benchmark in growing (exponentially) when lowering the temperature, whereas the $\alpha=1/2$ data saturate, in an unphysical fashion, at low $T$.

This procedure well exemplifies how the fluctuation diagnostics tools can serve as valuable source of information for refining the application of specific algorithms. This route of applications has been demonstrated to lead to significant improvements of results in the context of TRILEX. However, its potential applicability in this context is much more general. For instance, also in the framework of ladder approximations for parquet schemes (e.g., parquet-D$\Gamma$A vs. ladder-D$\Gamma$A) it can turn out immensely insightful to determine the leading fluctuation channel, before a specific (channel-biased) method is applied.

\section{Conclusions and outlook}

The rapidly increasing ability of computing one- and two-particle correlations functions of many-electron systems on an equal footing has allowed for the development of novel post-processing tools designed to quantitatively identify the physical mechanisms shaping the one-particle spectral properties, as well as the response functions.

Though sharing similar goals and -to some extent- philosophy, these novel approaches can be grouped in two main classes, depending on whether they are based (i) on a direct decomposition of the full vertex functions within the parquet formalism (``parquet decomposition" \cite{Gunnarsson2016}, heuristically associated to the motto {\it ``divide et impera"}, or rather (ii) on changes of representation of the Schwinger-Dyson equation for the self-energy (``fluctuation diagnostics" \cite{Gunnarsson2015}, under the motto {\it ``mutatis mutandis"}).

In both cases these post-processing schemes \cite{Rohringer2020} allow for quantifying the contribution of the different fluctuations (spin/charge/pairing) active, and sometimes competing,  in determining the main features of a given spectral (or, in some cases like \cite{Kauch2020}, response) function of many-electron systems.
From a more technical perspective,  we have discussed how the applicability of the first diagnostic tool gets typically restricted to the perturbative regime, because of the almost ubiquitous occurrence of divergences of 2PI vertices in many-electron problems \cite{Schaefer2013,Janis2014,Schaefer2016c,Gunnarsson2016,Vucicevic2018,Thunstrom2018,Chalupa2018,Springer2020}. 

While the newly introduced SBE \cite{Krien2019c} might allow to circumvent this issue in the near future preserving, at the same time, the high versatility of the parquet decomposition, the second diagnostic approach is already designed to be applicable in the whole parameter regime, independent on the divergences of irreducible vertices.

In the past few years, both classes of post-processing tools have been widely applied to ``diagnose" self-energy and response functions computed with the most advanced numerical techniques, ranging from the parquet \cite{yang2009,tam2013,Li2016,Wentzell2020,Eckhardt2020} and bosonic-exchange parquet \cite{Krien2020,Krien2020b} solvers, cluster \cite{Maier2005} and diagrammatic \cite{RMPVertex} extensions of DMFT and diagrammatic Monte Carlo \cite{Prokofev1998, Kozik2010, Rossi2017}. 
From the physical point of view, most applications have focused on the two-dimensional repulsive and attractive Hubbard model at different doping levels, allowing, among other results, a quantitative identification of the collective modes controlling the strong-momentum differentiation and the pseudogap spectral feature emerging in the hole-doped, low-$T$ regime of the two-dimensional Hubbard model. The latter could be essentially ascribed  \cite{Gunnarsson2015,Gunacker2016,Wu2016,Dong2020,Arzhang2020} to AF-spin fluctuation (slightly incommensurate for the nodal self-energy  \cite{Wu2016}), with a marginal role played by all other collective mode.

The quantitatively precise image obtained about such a hotly debated subject such as the origin of the pseudogap in the Hubbard model outlines the high potential of these new diagnostic tools for the post-processing of many-electron calculations.
We expect that this will provide a strong motivation for further extending these approaches in several directions, e.g., to make them applicable to multi-orbital \cite{Galler2016,Valenti2020,Kaufmann2020} or magnetically ordered \cite{DelRe2020} systems and/or also to the case of non-local electronic interactions.

\ack
We thank O. Gunnarsson, A. Kauch, E. Gull, O. Parcollet, M. Ferrero, J.P.F. LeBlanc, G. Rohringer, F. Krien, A. Valli, W. Metzner and especially A. Georges for interesting discussions and support. In addition we thank W. Metzner for carfully reading the manuscript. We thank U. Tozzi, R. Wien and S. Remo for moral support. The present work was supported by the Austrian Science Fund (FWF) through the Project I 2794-N35 (A.T.) and the Erwin-Schr\"odinger Fellowship J 4266 - ``{\sl Superconductivity in the vicinity of Mott insulators}'' (SuMo, T.S.). A.T. acknowledges the hospitality of the Coll{\`e}ge de France. We thank Regine Noack of the graphics service of the MPI-FKF Stuttgart for her graphical support.
\\\\
\bibliographystyle{unsrturl}
\bibliography{main}

\end{document}